\documentclass[]{aa}

\usepackage{graphicx,times,psfig}
\usepackage{graphics}

\begin{document}

   \title{The elemental abundances in the intracluster medium as
    observed with {\it XMM-Newton}}

   \author{
	T. Tamura \inst{1}
	\and
	J. S. Kaastra \inst{1}
	\and
	J.W.A. den. Herder \inst{1}
         \and
	J. A. M. Bleeker \inst{1}
	\and
	J. R. Peterson \inst{2}
	}

	   \offprints{T.Tamura}
	\mail{T.Tamura@plain.isas.ac.jp}

\institute{ SRON National Institute for Space Research, 
              Sorbonnelaan 2, 3584 CA Utrecht, The Nether\-lands 
	\and
	KIPAC, Stanford University
	PO Box 20450
	Stanford, CA 94309
}
   \date{Received ?; accepted 2004-3-2}

\abstract{
XMM-Newton observations of 19 galaxy clusters are used to measure the elemental abundances and their spatial distributions in the intracluster medium. 
The sample mainly consists of X-ray bright and relaxed clusters with a cD galaxy. 
Along with detailed Si, S and Fe radial abundance distributions within 300-700~kpc in radius, 
the O abundances are accurately derived in the central region of the clusters.
The Fe abundance maxima towards the cluster center, possibly due to the metals from the cD galaxy, are spatially resolved. 
The Si and S abundances also exhibit central increases in general, resulting in uniform Fe-Si-S ratios within the cluster.
In contrast, the O abundances are in general uniform over the cluster.
The mean O to Fe ratio within the cluster core is sub-solar, 
while that of the cluster scale is larger than the solar ratio.
These measurements indicate that most of the Fe-Si-S and O in the intracluster medium have different origins, presumably in supernovae Ia and II, respectively. 
The obtained Fe and O mass are also used to discuss the past star formation history in clusters.
}
\titlerunning{The ICM metallicity from {\it XMM-Newton}}
\maketitle

\section{Introduction}

Metals in the universe are being produced as a product of stellar evolution and
ejected into intergalactic space via stellar winds and supernovae.  
The location of these metals can be measured by X-ray observations of the intra-cluster medium
(ICM).  These observations indicate that the ICM contains an amount
of metals comparable to the total amount of metals found in galaxies 
(e.g. Tsuru \cite{tsuru91}; Arnaud et al. \cite{arnaud92}). 
Therefore, X-ray measurement of the ICM metallicity distribution is a unique way to study
the chemical history of galaxies.  Most importantly, the total amount of metals
and the abundance ratios among elements (e.g.  O/Fe ratio) constrain the total
number of stars that have formed as well as their origin (i.e.  the relative
ratios of supernova types; Renzini et al. \cite{renzini93}).

Measurements from
{\it ASCA} and {\it BeppoSAX} observations revealed important properties of the ICM metallicity.
These include an excess of Fe around cD galaxies (e.g. Fukazawa et al. \cite{fukazawa00}; De Grandi and Molendi \cite{grandi01}) and variations of the Si/Fe ratio within a cluster (e.g. Finoguenov et al. \cite{finoguenov00}) and among clusters (e.g. Fukazawa et al. \cite{fukazawa98}).
The variations in Si/Fe suggest that the metals in the ICM have at least two different origins, presumably SN Ia and SN II.
The spatial resolution of these previous
instruments was limited, causing the derived total amount of metals to depend on
their assumed spatial distribution.
The low spatial resolution combined with the limited spectral resolution also introduced 
a systematic uncertainty in the temperature structure. This is particularly important in the central cool regions.
This ambiguity in turn caused severe errors in the abundances of Fe and other elements.
Furthermore, in most cases these measurements are limited to the Fe, Si and S abundances.

Recent observations with {\it Chandra} and {\it XMM-Newton} have allowed a detailed
study of the ICM metals, using an improved spatial resolution and sensitivity
not only for the Si, S, and Fe lines but also for the O emission.  For example,
Tamura et al.  (\cite{tamura01b}) measured the spatial distribution of the O,
Si, S, and Fe in Abell~496 and found a difference between O and Si-S-Fe.  
David et al. (\cite{david01}) used the Chandra data of the Hydra-A cluster to measure the Si and Fe abundances.
In addition, the abundance distribution of several elements in the central region
(70~kpc in radius) of the Virgo cluster was obtained with high accuracy
(Finoguenov et al.  \cite{finoguenov02}; Gasteldello \& Molendi \cite{gas02};
Matsushita et al.  \cite{matsushita03}).

In this paper, we use the {\it XMM-Newton} EPIC and RGS instruments and measure
the spatial distribution of heavy elements in 19 clusters.  
This systematic analysis enables us to compare the metal distribution in the sample. 
In addition, we may reduce the statistical errors on any given measurement when the variation in the sample is negligible.
We selected the sample based on their X-ray flux and spatial size. 
Therefore the sample is limited to nearby X-ray bright clusters
with a mean redshift of 0.06.

We obtained the radial distribution of Fe on a sub-arcmin scale for each cluster
and confirmed the central increase of the Fe abundance within a typical radius
of 100~kpc in most cases.  Furthermore, radial profiles of O, Si, and S relative
to Fe have been measured in a fair number of clusters.  We found that the O and
Si-S-Fe abundances have different spatial distributions when we combine the results from
several clusters.

Throughout this paper, we assume the Hubble constant to be $H_0 = 100h$
km\,s$^{-1}$\,Mpc$^{-1}$ and use 90\% confidence levels unless stated otherwise.
Solar abundances are taken from Anders \& Grevesse (\cite{anders}) with
Fe/H=$4.68\times 10^{-5}$.

\section{Observations}

\begin{table*}[!ht]
\caption{Cluster sample.}
\label{tbl:sample}
\begin{center}
\begin{tabular}{lrrrrrrrrrr}
\hline
\multicolumn{1}{c}{Cluster} & 
\multicolumn{1}{c}{redshift $^{\mathrm{a}}$} & 
\multicolumn{1}{c}{$kT$ $^{\mathrm{b}}$} & 
\multicolumn{1}{c}{$n_{\rm H}$ $^{\mathrm{c}}$} & 
\multicolumn{1}{c}{C $^{\mathrm{d}}$} & 
\multicolumn{1}{c}{XMM} &
\multicolumn{3}{c}{Exposure time (ks)$^{\mathrm{e}}$} & 
\multicolumn{2}{c}{R$^{\mathrm{f}}$} \\
 & & (keV) & ($10^{24}$~m$^{-2}$)& & revolution & MOS & pn & RGS & IN  & OUT\\
\hline
NGC 533                & 0.0175 & 1.3 &  3.00  &	C	&  195 & 38 & 31  &  49 & 0 & 6\\
A 262                  & 0.0155 & 2.2 &  8.94 (5.33)&	C	&  203 & 24 & 15  &  35 & 0 & 12\\
S\'ersic~159$-$03      & 0.0572 & 2.4 &  1.79  &	C/SE	&   77 & 31 & 28  &  36 & 0 & 9 \\
MKW 9                  & 0.0402 & 2.6 &  4.18  &	C	&  311 & 22 & 13  &  -- & 0 & 3\\
2A 0335+096            & 0.0344 & 3.0 & 28.71 (18.64)&	M	&  215 & 11 &  5  &  18 & 0 & 9\\
A 2052                 & 0.0356 & 3.1 &  2.91  &	M/SE	&  128 & 30 & 23  &  28 & 0 & 12\\
Hydra A  (A 780)       & 0.0550 & 3.4 &  4.80  &	M	&  183 & 17 &  3  &  25 & 0 & 6\\
MKW 3s                 & 0.0455 & 3.5 &  2.89  &	M/SE	&  129 & 37 & 32  &  -- & 0 & 12\\
A 4059                 & 0.0466 & 4.0 &  1.06  &	M	&  176 & 34 &  6  &  44 & 0 & 9\\
A 1837                 & 0.0707 & 4.4 &  4.38  & 	M	&  200 & 48 & 40 &  -- & 0 & 6\\
A 496                  & 0.0322 & 4.4 &  6.44 (4.23) &	M	&  211 &  9 &  5  &  18 & 0 & 12\\
A 3112                 & 0.0756 & 4.5 &  2.61  &	M/SE	&  191 & 22 & 16  &  24 & 0 & 9\\
A 1795                 & 0.0639 & 5.8 &  1.01  &	M/SE	&  100 & 29 & 22  &  37 & 0 & 9\\
A 399                  & 0.0706 & 6.2 & 10.90  &	H	&  127 & 13 &  6  &  -- & 0.5 & 9\\
Perseus (A 426)        & 0.0179 & 6.5 & 14.90  &	H	&  210 & 43 & 33  &  -- & 0 & 12\\
A 1835                 & 0.2541 & 7.2 &  2.32  &	H	&  101 &  0 & 25  &  29 & 0 & 6 \\
Coma (A 1656)          & 0.0240 & 7.5 &  0.89  &	H/SE	&   86 & 15 & 13  &  -- & 0.5 & 12\\
A 754                  & 0.0561 & 8.0 &  5.67 (4.37) &	H	&  262 & 14 & 11  &  -- & 0.5 & 12\\
A 3266                 & 0.0614 & 8.7 &  1.60  &	H	&  153 & 24 & 19  &  -- & 0.5 & 12\\
\hline
Virgo$^{\mathrm{g}}$  & 0.0027 & 2.6 &  1.80  &	C	&   97 & 34 & 25  &  -- & 0 & 12\\
\hline
\end{tabular}
\begin{list}{}{}
\item[$^{\mathrm{a}}$] Redshift, to be used for the distance estimate.
\item[$^{\mathrm{b}}$] The temperature outside the central cool region obtained from 
                       our EPIC analysis (Kaastra et al. \cite{kaastra04}).
\item[$^{\mathrm{c}}$] Galactic column density to be used for the spectral modeling.
In most cases, these are obtained from \ion{H}{i} map (Dickey \& Lockman \cite{dickey}).
In other cases, we present the used (X-ray) value with the \ion{H}{i} 
value in parentheses.
\item[$^{\mathrm{d}}$] Classification of clusters used in the present paper.
``C'', ``M'', and ``H'' indicate cool, medium-temperature, and hot clusters, respectively.
``SE'' indicates clusters which show soft X-ray excess.
\item[$^{\mathrm{e}}$] Net exposure time, after rejection of high background time periods.
We did not use the RGS data for clusters indicated by ``--'', 
because of poor statistics. 
\item[$^{\mathrm{f}}$] Innermost and outermost radii for the EPIC spectral analysis in arcmin.
\item[$^{\mathrm{g}}$] The data for the Virgo cluster are used only for the EPIC calibration. 
\end{list}
\end{center}
\end{table*}

We analyzed 19 clusters taken from GT and PV observations of {\it
XMM-Newton} (Table~\ref{tbl:sample}).  The RGS exposure times are sometimes
shorter than those reported in Peterson et al.  (\cite{peterson03}), because of
a difference in the criteria for rejecting the exposure period with high background.
We have previously reported detailed
results for A~1795 (Tamura et al.  \cite{tamura01a}), S\'ersic~159$-$03 (Kaastra
et al.  \cite{kaastra01}), A~1835 (Peterson et al.  \cite{peterson01}), and
A~496 (Tamura et al.  \cite{tamura01b}).

The present paper is one of a series of papers with a systematic study of
cluster properties using similar data sets:  soft X-ray excess (Kaastra et al.
\cite{kaastra03a}); central cooling region using RGS data (Peterson et al.
\cite{peterson03}); central cooling using EPIC (Kaastra et al.
\cite{kaastra04}).  Here we focus on the spatial distribution of the metal
abundances using both the EPIC and RGS data.

A detailed description of the {\it XMM-Newton} satellite and its instruments is
found in Jansen et al.  (\cite{jansen01}), Turner et al. (EPIC/MOS; \cite{turner01}),
Str\"uder et al. (EPIC/pn; \cite{struder01}), and den Herder et al.  (RGS;
\cite{herder01}).

\section{The EPIC analysis}

\subsection{Spectral extraction}
In order to derive the spatial metallicity distribution, we used
spatially-resolved EPIC spectra.  We used the Science Analysis System (SAS;
version 5.3.3) for the basic data processing.  
Then, we corrected the filtered events for the background, the telescope point spread function (PSF), 
and the effective area.
We described the data correction and extraction extensively in Kaastra et al. (\cite{kaastra04}).  

We removed high background time periods based on the number of counts in the 10--12 keV band over the full field of view.
Even after rejecting these high background events, there were still variations in the instrumental background from observation to observation.
In order to correct for thiso weak variation, we sorted the cluster and blank sky data into subsets with the same 10--12~keV count rate.
Then, the frequency distributions of source count rate and a set of the blank sky spectra at different 10--12~keV count rate level were used to determine the subtracted background spectrum.
The source and background spectra were obtained from the same detector regions.
To compensate for remaining variations in the instrumental background, 
we included a systematic uncertainty of at least 10~\% in the spectral fitting.

In both MOS and pn the Al lines around 1.48~keV, which are close to the line positions of H-like and He-like Mg ions, are strong.
Therefore, we did not attempt to derive the Mg abundance except for bright cluster central regions.
In addition, in the MOS, there are instrumental Si lines around 1.74~keV.
Compared to the Al lines, these lines are relatively weak and not as close to that from astrophysical Si ions.
Therefore, an additional uncertainty in the Si background lines produces no change in our result.
We verified this using representative cluster data.

In the low energy band, the cosmic background dominates the instrumental background and varies spatially by more than 10\%.
Therefore, we included larger systematic errors in the soft X-ray band as shown in Table~\ref{tbl:sys}.

In all observations the bright X-ray point sources were removed. 
Typically, a median of 8 point sources (for MOS1 and MOS2) or 15 (for pn) were discarded.

Then, we made a background subtracted radial profile for each energy.
This profile was corrected for the exposure, telescope PSF, and vignetting.
We assumed energy-independent PSF correction (Kaastra et al. \cite{kaastra04}).
Therefore it affects only the spectral normalization, not its shape.
On the other hand, 
the correction for the vignetting due to the telescope is dependent on the X-ray position and energy.
The MOS cameras also have a correction due to obscuration of the Reflection Grating
Array that is dependent on position and energy as well.

From the corrected radial profile for each energy band, 
we extracted {\it projected} spectra from concentric annuli centered on the
X-ray maximum with outer radii of 0.5, 1, 2, 3, 4, 6, 9, 12, 16 arcminutes.  
In addition, assuming spherical symmetry, we derived {\it deprojected} spectra from shells with the same radial binning as above.
The deprojected spectrum represents a set of count rates in a spherical shell between the two spheres and can be given 
by a linear combination of the observed number of counts in the (projected) surface brightness profile.
Note that we ignored X-ray emission outside the field of view 
and that the spectra of the outermost annulus (12\arcmin--16\arcmin) are used only for 
the deprojection for the inner shells.
We used different innermost and outermost bins among our clusters as shown in
Table~\ref{tbl:sample} depending upon the statistical quality of the spectra.

We do not completely take into account the spectral mixing due to the PSF.
Therefore the true profiles of temperature and metallicity could be much steeper than the present results. 
This effect should be small since the the radial bin size is larger than the PSF width (a half energy width of $15$\arcsec).

\subsection{Spectral fitting}
In most cases, the instrumental background dominates the source emission above
$\sim 8$~keV.  Between 7--9~keV, the pn instrumental background
contains strong spectral lines.  
Furthermore, when we fitted the pn spectra of bright clusters with representative models, we found systematic residuals below the Fe-L lines.
 This is probably due to uncertainties in the energy redistribution response at the low energy.
Taking this into account, we restricted our analysis for the
MOS and pn spectra between 0.3 and 8~keV and between 0.7 and 7.0~keV energy ranges,
respectively.  The three spectra (MOS1, MOS2, and pn) were fitted simultaneously
with the same spectral model, but we allowed the relative normalizations to be
free parameters.


\begin{table}[!ht]
\caption{Systematic errors relative to the flux. These were
used for the EPIC spectral modeling and error estimates.}
\label{tbl:sys}
\begin{tabular}{ccccc}
\hline
Energy band (keV)	& 0.3--0.5 & 0.5--0.7 & 0.7--2.0 & 2.0--8.0\\
\hline
source 	      & 5	   & 5	      & 5	& 10\\
background    & 35	   & 25	      & 15	& 10\\
\hline
\end{tabular}
\end{table}


We used a 
collisional ionization equilibrium model (CIE; implemented in the SPEX package\footnote{see http://www.sron.nl/divisions/hea/spex/}
; Mewe et al. \cite{mewe85}; Kaastra \cite{kaastra92}; Liedahl et al. \cite{liedahl95})
modified by photoelectric
absorption.  
The abundance ratios of C and N were fixed to the solar values, 
based on the best-fit value derived from the RGS in the central region of M87 (Sakelliou et al. \cite{sakelliou02}).
Those of O, Ne, Mg, Si, S, Ar, Ca, and Fe were left free, while those
of other heavy elements were fixed to that of Fe (e.g., Ni/Fe=1).  Thus the free
parameters are the normalization, temperature ($kT$), and metal abundances (O,
Ne, Mg, Si, S, Ar, Ca, and Fe).

We used a fixed column density for each cluster as shown in
Table~\ref{tbl:sample}.  These column densities have been adjusted to match the
observed EPIC spectra as follows.  In most cases, the best-fit value is
consistent with the Galactic one obtained from the \ion{H}{i} map (Dickey \& Lockman \cite{dickey}).  In those
cases we kept the column density fixed to the Galactic value.  In A~262,
2A~0335+096, A~496 and A~754, however, the data require additional absorption,
possibly caused by Galactic gas and dust (we investigated these individual cases
in more detail in Kaastra et al.  \cite{kaastra03a}).


Usually, the best-fit parameters (normalization, $kT$, and abundances) are
derived at the same time by fitting the spectrum in a wide energy band.  
If however the response matrix is not accurate due to a calibration error in some
wide energy band of the spectrum, the modeled local continuum may differ
slightly from the actually observed continuum in that wide energy band.
This could result in errors on the derived equivalent width of lines in that band and
hence in errors on the obtained abundances.  
In fact, when we used this global fitting method with the MOS and pn spectra separately with bright clusters,
we found that the MOS gives systematically higher Si and S abundances by $\sim 0.2$ of the solar value than the pn does.
To avoid such errors as much as possible, 
we derive the parameters in two steps.  First, we measure the best-fit $kT$ by fitting the
wide energy band spectra.  Secondly, we fix $kT$, use the local spectra in the
limited energy band close to the relevant elemental-ion lines and derive the
abundance by allowing the normalization to be free.  We used energy bands of
0.35--1.0 (for O), 0.33--2.2 (Ne and Mg), 1.2--3.2 (Si and S), 2.7--4.7 (Ar and
Ca), all in keV.  An exception is the Fe abundance, for which we use the wide
energy band to include both the Fe-L and Fe-K lines simultaneously.


We analyzed both projected and deprojected spectra.  The projected spectra in
general have better statistics but poorer fits for each radius, mainly because
of the contamination from different radii with different $kT$.  Also, the
derived spatial distribution of the temperature changes by using
deprojection, in particular in the inner regions (See Fig.2 in Tamura et al. \cite{tamura01b} for an example).  
To obtain an accurate abundance distribution, proper modeling of the temperature distribution is necessary.
Therefore, we present the results of the deprojected spectra.

\subsection{Systematic uncertainties}
We used several calibration sources to estimate the uncertainty in the response
calibration.  Based on this, we place conservatively systematic errors on the
source spectra.  
Table~\ref{tbl:sys} summarizes the values of these systematic errors that
we used.
These errors along with those of the background were added in quadrature in the binned spectra.

\begin{figure*}[!ht]
\begin{center}
	\resizebox{0.48\hsize}{!}{\includegraphics[angle=-90]{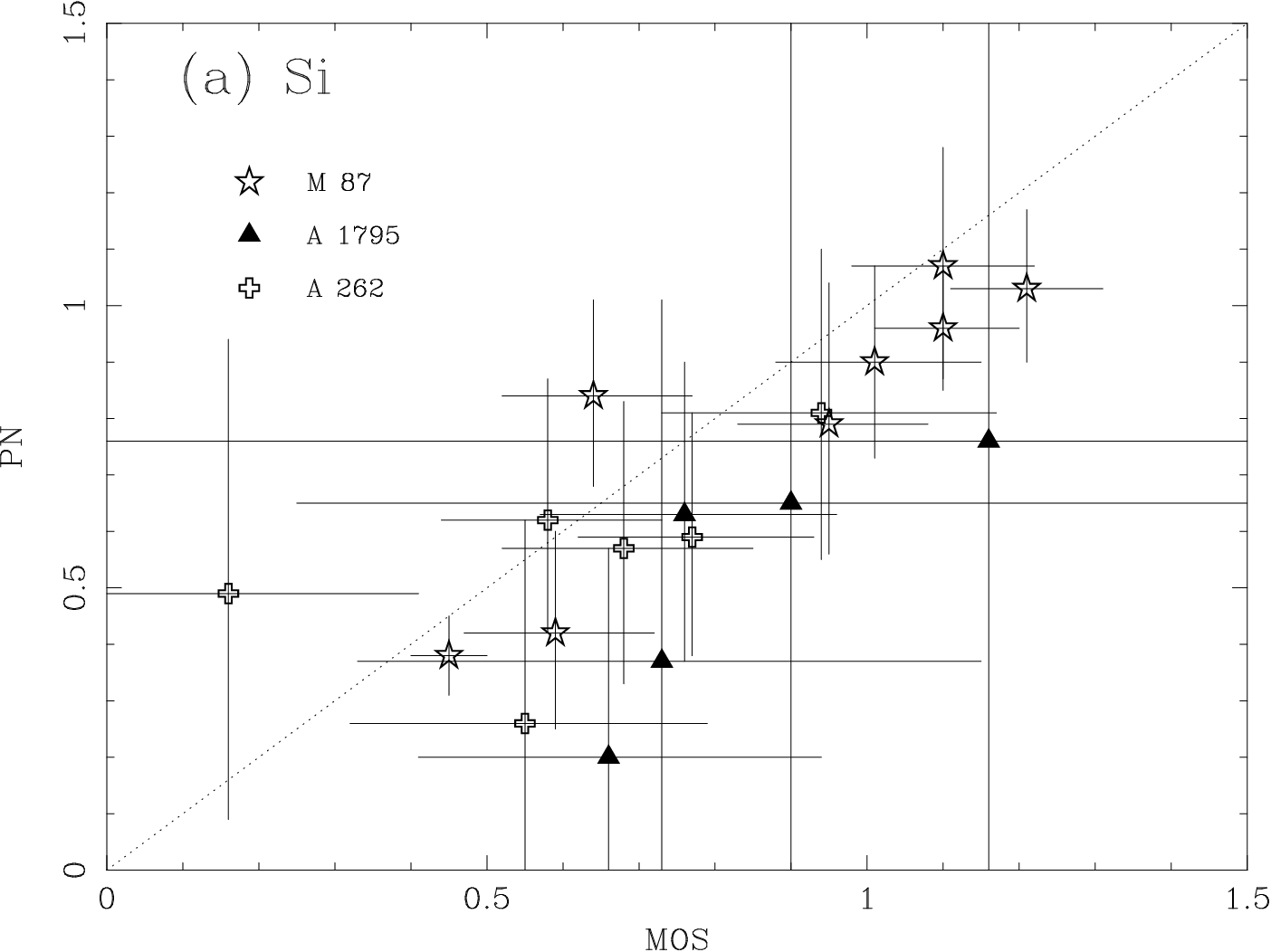}}
	\resizebox{0.48\hsize}{!}{\includegraphics[angle=-90]{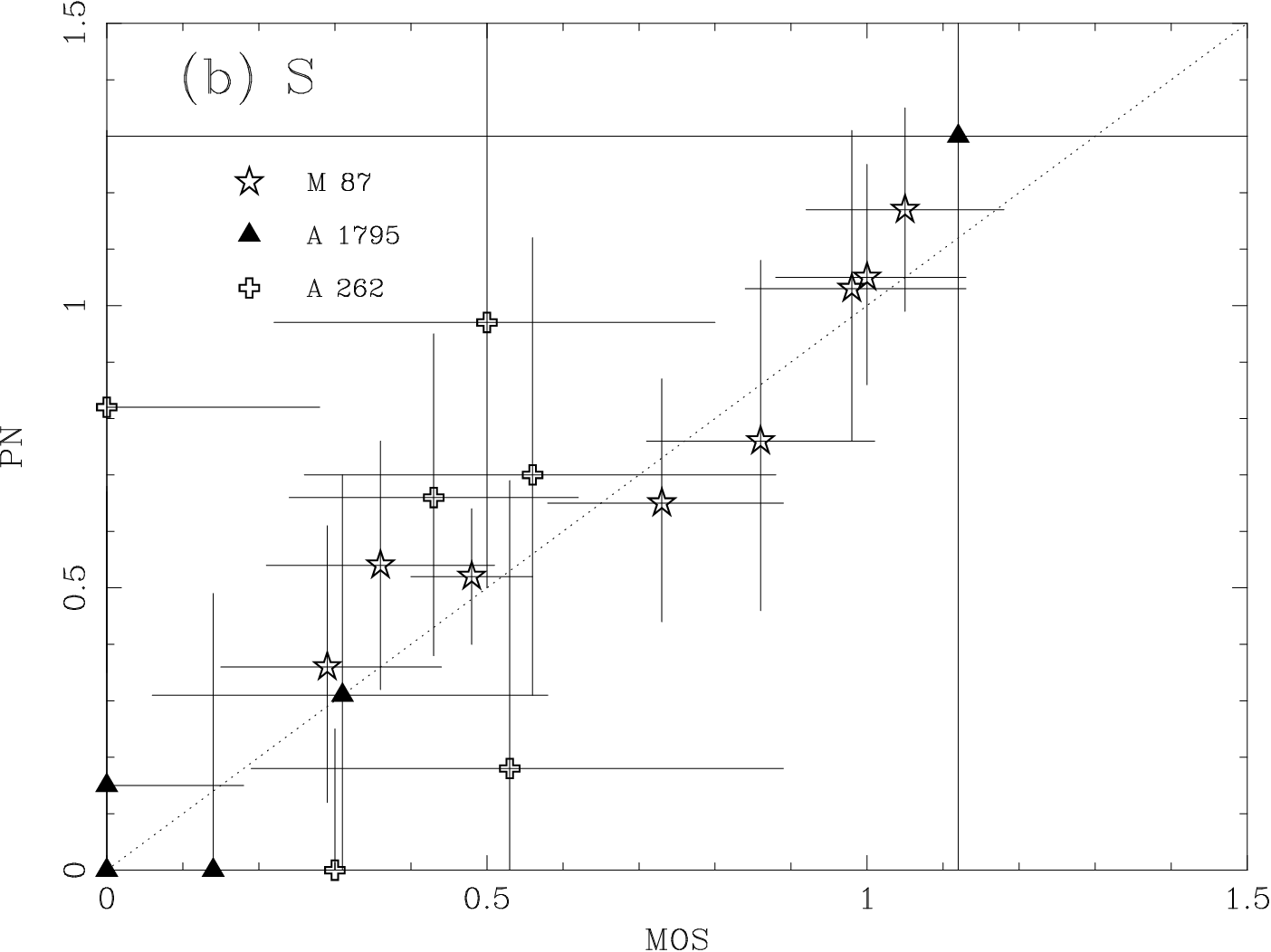}}
	\resizebox{0.48\hsize}{!}{\includegraphics[angle=-90]{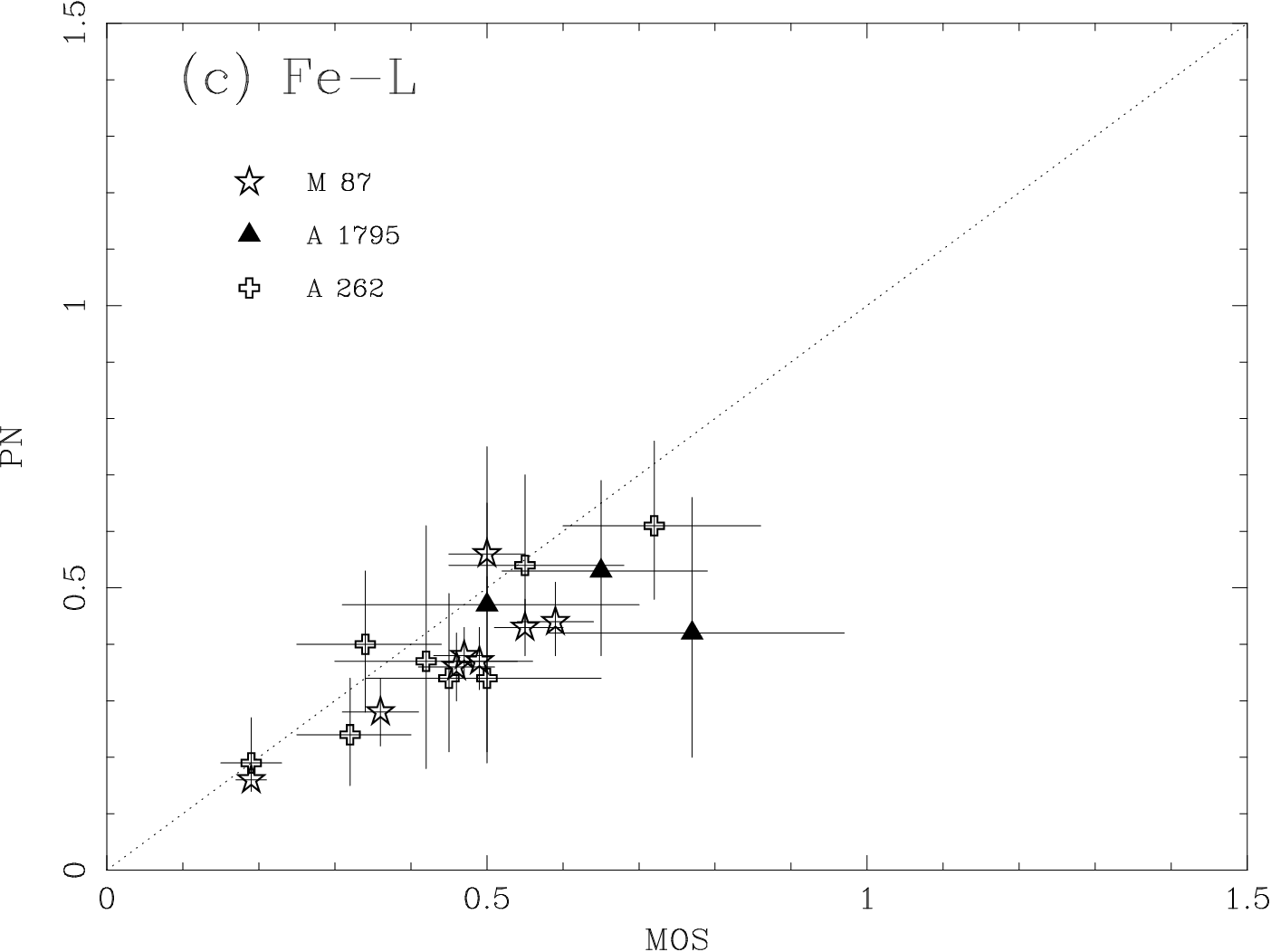}}
	\resizebox{0.48\hsize}{!}{\includegraphics[angle=-90]{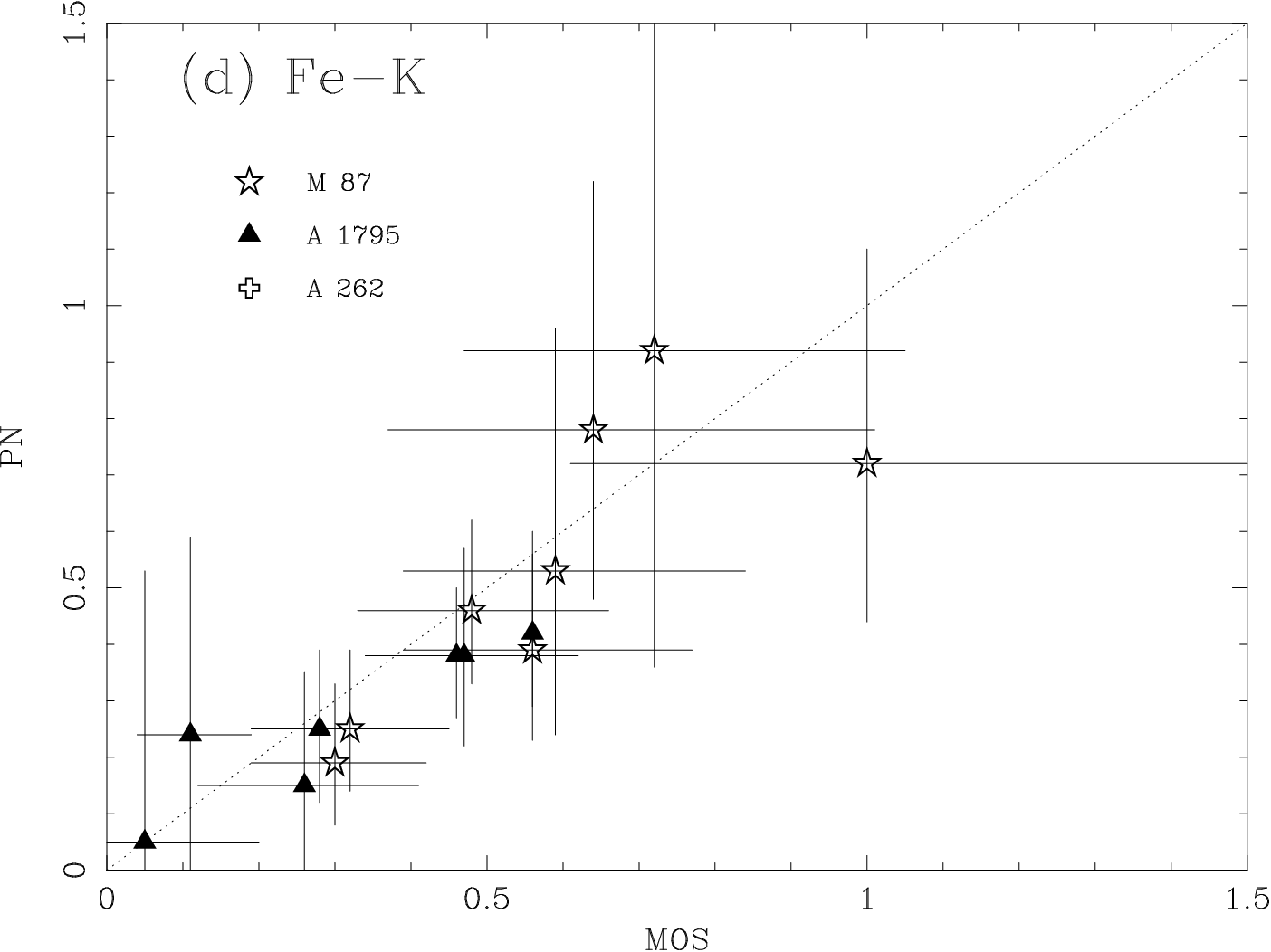}}
\caption{Comparison between the MOS and pn results for the Si (a), S (b), 
and Fe (c and d) abundances. 
We used spectra from several regions of M~87 (the brightest cluster at low energies), 
A~262 (temperature 2~keV), and A~1795 (temperature 4--6~keV) 
as representative samples.}
\label{fig:mos-pn}
\end{center}
\end{figure*}

\begin{figure}[!ht]
\begin{center}
\resizebox{\hsize}{!}{\includegraphics[angle=-90]{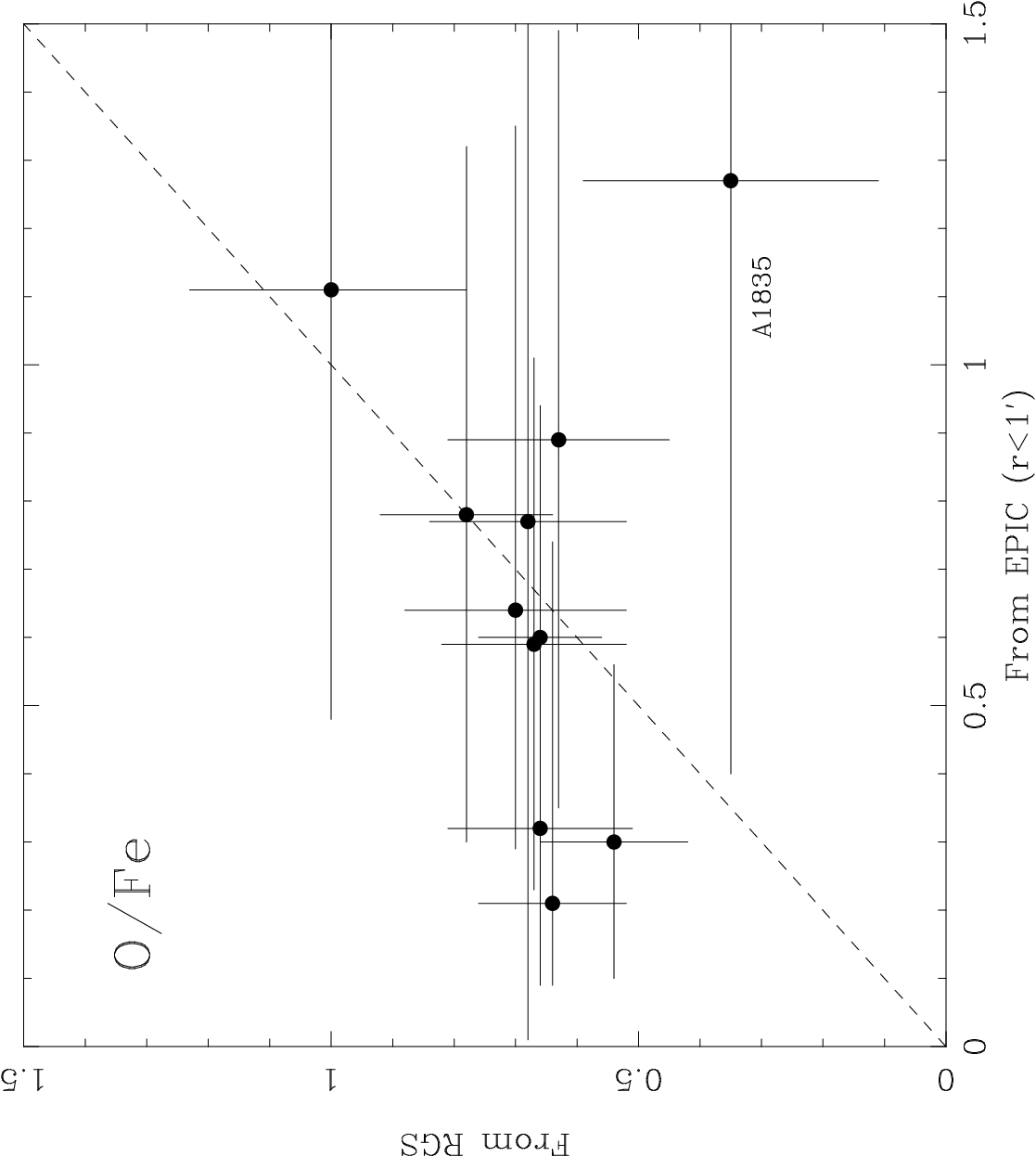}}
\resizebox{\hsize}{!}{\includegraphics[angle=-90]{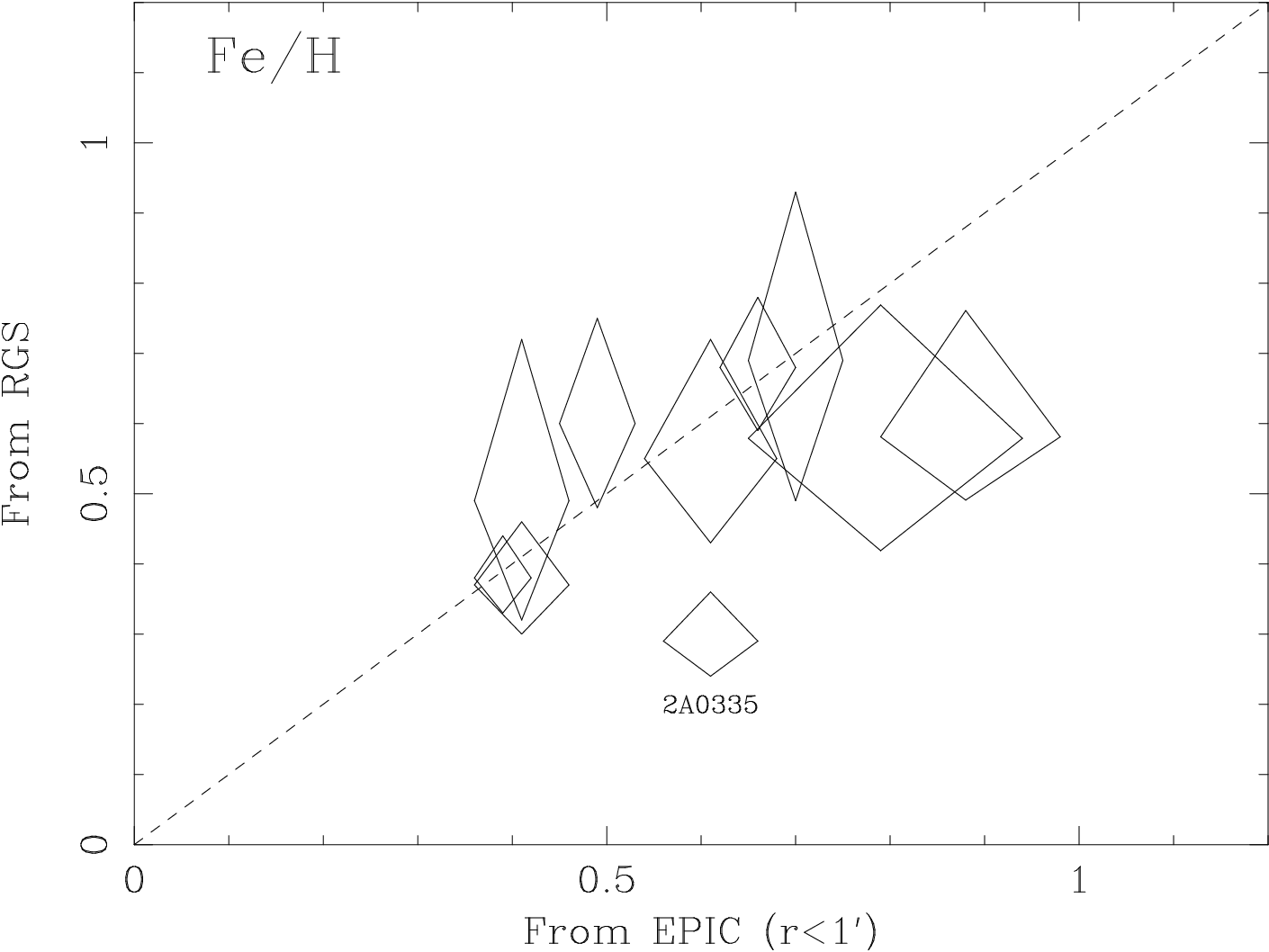}}
\caption{Comparison between the EPIC and RGS results for O/Fe and Fe/H.}
\label{fig:epic_vs_rgs}
\end{center}
\end{figure}

To assess additional systematic errors, we derived abundances (Si, S, and Fe)
separately from the MOS and pn spectra in representative clusters (M~87, A~262,
and A~1795).  For the Fe abundance determination, we separately used Fe-L and
Fe-K lines.  The results are shown in Fig.~\ref{fig:mos-pn}.  The pn detector
still gave systematically smaller Si and Fe abundances than MOS.  
The differences were less than 10--20~\% of the obtained values and less than the typical statistical
errors for each spectrum.  In other cases, the two instruments provided
consistent results.  Because the \ion{O}{viii} Ly$\alpha$ line was contained
only in the MOS spectrum in our case, we cannot check the O abundance in a
similar way.  Nevertheless, we confirmed that the O abundances obtained from MOS
are consistent with those obtained from RGS as shown below
(Fig.~\ref{fig:epic_vs_rgs}).  The three types of {\it XMM-Newton} instruments have
different energy responses and have been calibrated independently, except for the telescope response.
Therefore the above consistency checks indicate that there is 
negligible additional systematic error above the values that we already
assigned compared to the statistical uncertainty in our sample.

\subsection{Results}


\begin{figure}[!ht]
\resizebox{0.7\hsize}{!}{\includegraphics[angle=-90]{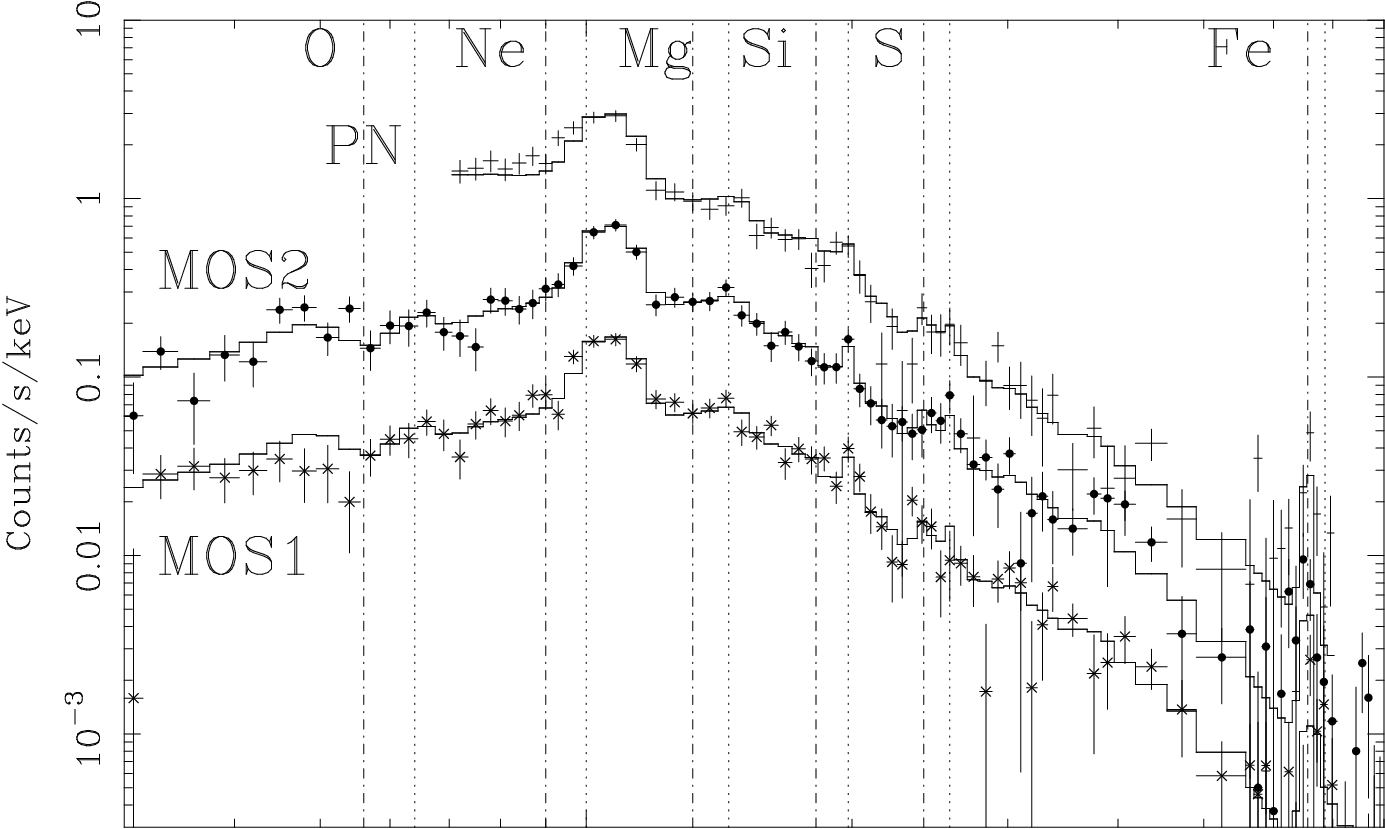}}
\resizebox{0.7\hsize}{!}{\includegraphics[angle=-90]{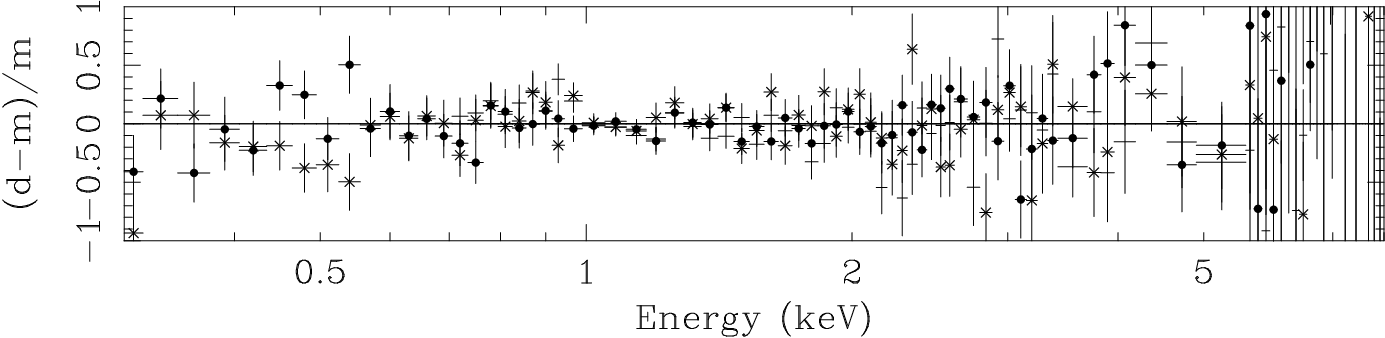}}
\caption{An example of the spectral fitting.  The spectra of A~262 between
1\arcmin--2\arcmin\ in deprojected radius are shown.  
The MOS1 and pn spectra are multiplied by factors of 0.25 and 2, 
respectively, for clarity. The data are shown with
the model as a histogram.  Expected line positions for H- and He-like
ions are indicated by vertical lines.  
The bottom panel shows the fit residuals [(data--model)/model].
In this case, we obtain a temperature of
1.9~keV and abundances for O, Si, S, and Fe of 0.5, 1.0, 1.0, and 0.7~times
solar, respectively.}
\label{fig:a262}
\end{figure}

\begin{figure}[!ht]
\resizebox{0.7\hsize}{!}{\includegraphics[angle=-90]{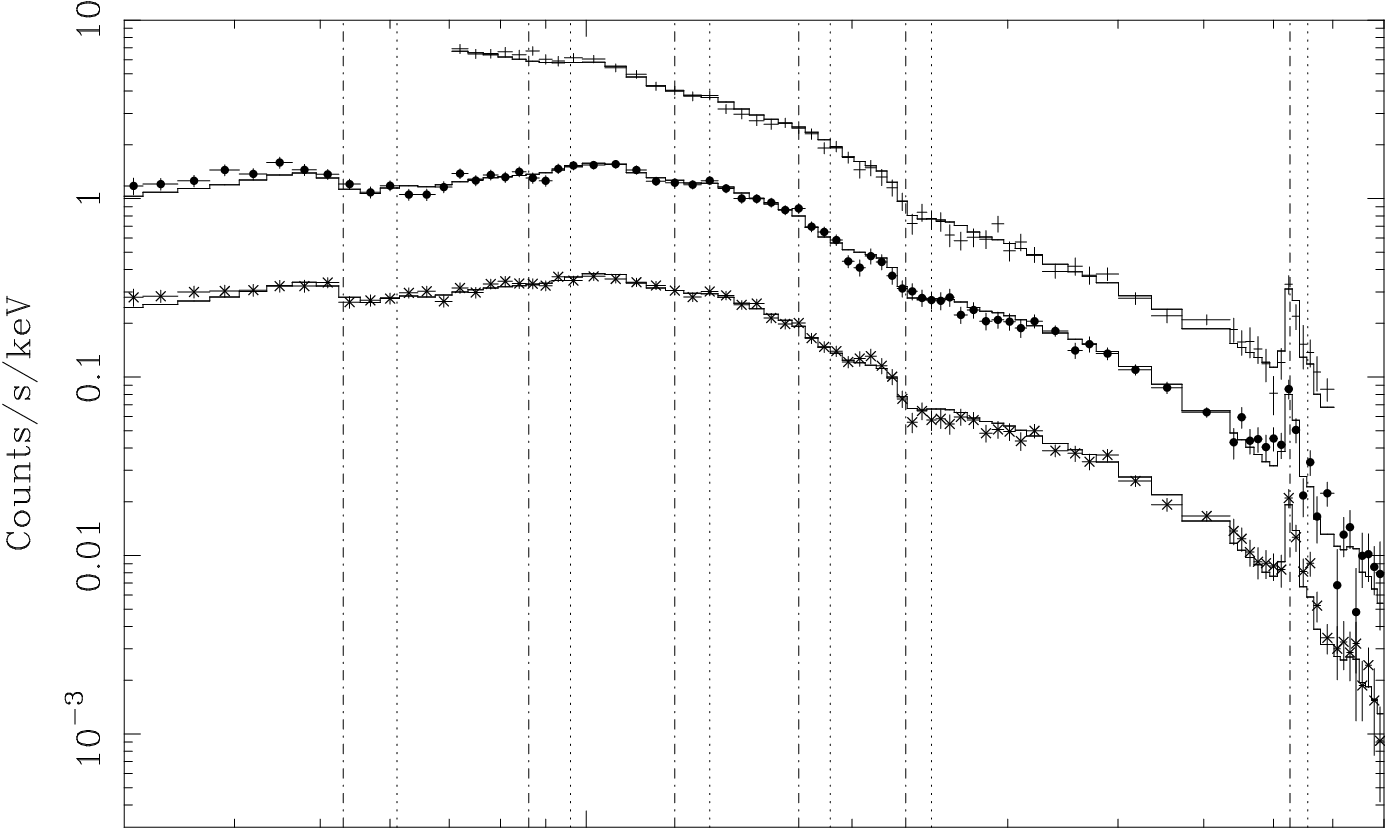}}
\resizebox{0.7\hsize}{!}{\includegraphics[angle=-90]{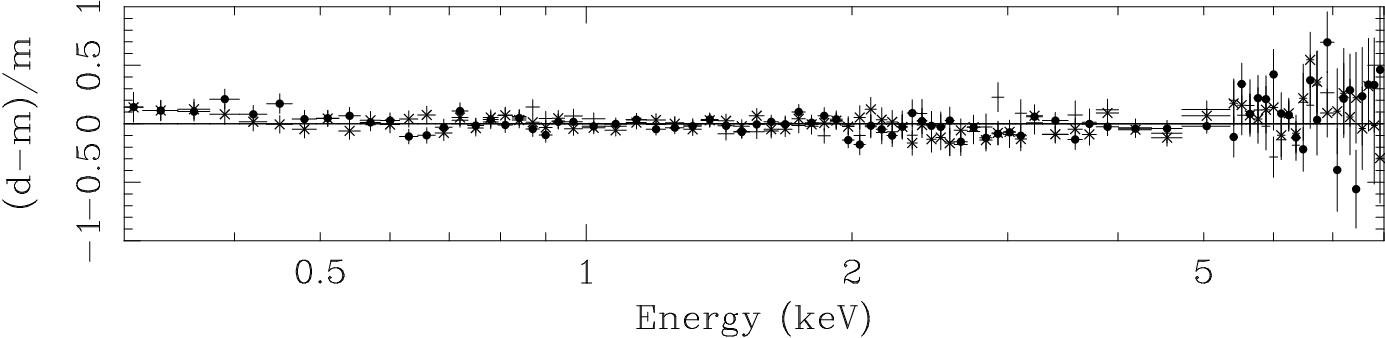}}
\caption{Same as the previous figure, but for the spectra of A~1795
between 1\arcmin--2\arcmin\ in deprojected radius.
We obtain  a temperature of 5.3~keV and abundances for
O, Si, S, and Fe  of  0.20, 0.7, 0.0, and 0.4~times solar, respectively.}
\label{fig:a1795}
\end{figure}

In general, the single temperature (1T) model provided a good fit to the
spectra.  Figures~\ref{fig:a262} and \ref{fig:a1795} show examples of the fit.
However, the spectra from the following regions cannot be described by the 1T model ($\chi^2/\nu>1.2$).
NGC~533 ($0'-1'.0$), 
A~262 ($0'-0'.5$),
2A~0335+096 ($0'-1'.0$),
MKW~3s ($6'-12'$),
A~2052 ($0'-0'.5, 9'-12'$), 
A~4059 ($0'-0'.5$), 
A~3112 ($1'-2', 3'-6'$), 
and Perseus ($0'-2', 3'-4'$).

In the inner regions of some clusters
($r<$1\arcmin), the data exhibit a clear deviation from the 1T model.  A major
part of this discrepancy can be removed by adding an additional cool component.
In addition, some spectra in the outer regions cannot be modeled by the 1T
model, probably because of soft excess emission below $0.5$~keV.  We reported
earlier a detailed analysis of this soft excess in S\'ersic~159$-$03, 
A~2052, A~1795, and Coma (Kaastra et al.  \cite{kaastra03a}).  In addition to
these clusters, A~3112 also shows a similar excess above the 1T model
(Nevalainen et al.  \cite{nevalainen03}; Kaastra et al.  \cite{kaastra03c}).


To describe the multi-temperature structure in the central regions, we used a
two temperature model (2T).  In Kaastra et al.  (\cite{kaastra04}) and Peterson
et al.  (\cite{peterson03}) we found that in almost all cases, at each radius
there is negligible emission from gas with a temperature less than one-third to
half of the ambient temperature.  Therefore, we fixed the temperature separation
between both components as $T_0 = 2 T_1$, where $T_0$ and $T_1$ are the two
temperatures.

\begin{table*}[!ht]
\begin{center}
\caption{Spectral fit results for the inner regions of some clusters
for a single temperature (1T fit) and a two temperature (2T fit) model.
We present no error when the fit is largely unacceptable.}
\label{tbl:in-fit}
\begin{tabular}{lccccccc}
\hline
cluster	& R$^{\mathrm{a}}$ & \multicolumn{3}{c}{1T fit} & \multicolumn{3}{c}{2T fit} \\
	& 	   & $kT^{\mathrm{b}}$	& Fe/H  & $\chi^2/\nu$ & $kT_{0}^{\mathrm{b}}$ & Fe/H & $\chi^2/\nu$ \\
	& 	   & (keV) & (solar) & & (keV) & (solar) & \\ 
\hline
A~262	    & 0.0--0.5 & 1.0 & 0.31 & 457/227 & $1.5\pm0.07$ & $0.85\pm0.1$ & 274/226 \\
   	    & 0.5--1.0 & $1.4\pm0.04$ & $0.58\pm0.07$ & 242/227 & $1.6\pm0.1$ & $0.89\pm0.1$ & 214/226 \\
2A~0335+096 & 0.0--0.5 & 1.4 & 0.32 & 400/228 & $1.7\pm0.1$ & $0.55\pm 0.07$ & 298/227 \\
   	    & 0.5--1.0 & $1.8\pm0.05$ & $0.5\pm0.05$  & 275/227 & $2.1\pm0.1$ & $0.64\pm 0.06$ & 210/227 \\
A~2052	    & 0.0--0.5 & 1.4 & 0.35 & 574/228 & $2.0\pm0.1$ & $0.62\pm0.05$ & 312/227\\
   	    & 0.5--1.0 & $2.6\pm0.06$ & $0.71\pm0.06$ & 255/228 & $2.9\pm0.2$ & $0.67\pm0.06$ & 213/227\\
A~4059	    & 0.0--0.5 & 2.2 & 0.63 & 283/226 & $2.6\pm 0.4$ & $0.60\pm0.2$ & 258/223\\
\hline
\end{tabular}
\begin{list}{}{}
\item[$^{\mathrm{a}}$] Radial range (arcmin).
\item[$^{\mathrm{b}}$] The best-fit and hotter temperature for the 1T fit and for the 2T fit, respectively.

\end{list}
\end{center}
\end{table*}

For the inner regions of A~262, 2A~0335+096, A~2052 and A~4059, the 2T model
improved the fit (Table~\ref{tbl:in-fit}).  In some cases, the Fe abundance
changed significantly as compared to the single temperature fit (the so-called
Fe-bias, see for example Buote \cite{buote00} and Molendi \& Gastaldello
\cite{molendi01}).  For these clusters, we use the abundances derived from this
2T model for the discussion below.  For the inner regions in some other clusters
(e.g.  A~496 and A~1795), the 2T model sometimes provides a slightly better fit
over the 1T model.  However, in these cases the changes in the obtained Fe
abundances are small compared to the statistical errors.  This is because the
deviation from isothermality within each radius is relatively small.

In some cases, the fit is improved further by de-coupling $T_0$ with $T_1$.
However, even in those cases, 
the change in the Fe abundance from the above 2T model is small ($\sim $10\%) compared to the errors.

\begin{figure*}[!ht]
\begin{center}
	\resizebox{0.45\hsize}{!}{\includegraphics[]{rkpc_fe.qdp.vps}}
	\resizebox{0.45\hsize}{!}{\includegraphics[]{rkpc_o2fe_rebin.qdp.vps}}
	\resizebox{0.45\hsize}{!}{\includegraphics[]{rkpc_si2fe_rebin.qdp.vps}}
	\resizebox{0.45\hsize}{!}{\includegraphics[]{rkpc_s2fe_rebin.qdp.vps}}
\caption{Radial profiles of the Fe/H, O/Fe, Si/Fe and S/Fe ratios in solar
units, derived from the EPIC spectra.  The best-fit values for cool,
medium-temperature, hot, and soft excess clusters are shown with open-circles,
filled-circles, open-square, and ``$\times$'', respectively.  Typical error bars
are shown for A~262, A~496, and Perseus.  Mean values and errors for the cool
and medium-temperature clusters (T$<6$~keV), except for the soft excess
clusters, are shown with diamonds.  In panel (b), the O/Fe ratio obtained from
the RGS spectra are also shown with filled-stars.}
\label{fig:epic-radial-ab}
\end{center}
\end{figure*}

\begin{table}[!ht]
\begin{center}
\caption{Abundances (scaled to the solar ratio) obtained from the EPIC analysis for each cluster. 
Values with errors larger than 3 times solar are not presented 
(indicated by $-$).}
\label{tbl:eack-abun}
\begin{tabular}{rrrrr}
\hline 
R$^{\mathrm{a}}$ & Fe/H& O/Fe& Si/Fe & S/Fe\\
\hline 
\multicolumn{5}{c}{NGC~533} \\ 
\#0  & $0.39\pm {0.06}$  & $ 0.3\pm { 0.7}$  & $ 1.6\pm { 0.7}$  & $ 2.8\pm { 1.3}$ \\ 
\#1  & $0.26\pm {0.09}$  & $ 1.9\pm { 1.9}$  & $ 1.7\pm { 1.2}$  & $ 1.1\pm { 1.7}$ \\ 
\hline 
\multicolumn{5}{c}{A 262} \\ 
\#0  & $0.67\pm {0.05}$  & $ 0.6\pm { 0.3}$  & $ 1.4\pm { 0.2}$  & $ 1.0\pm { 0.3}$ \\ 
\#1  & $0.34\pm {0.04}$  & $ 1.8\pm { 0.9}$  & $ 1.2\pm { 0.4}$  & $ 0.7\pm { 0.5}$ \\ 
\hline 
\multicolumn{5}{c}{A 1837} \\ 
\#0  & $0.64\pm {0.14}$  & $ 0.5\pm { 1.6}$  & $ 2.8\pm { 1.6}$  & $ 1.0\pm { 1.4}$ \\ 
\#1  & $0.39\pm {0.07}$  & $ 0.9\pm { 1.1}$  & $ 0.9\pm { 0.7}$  & $ 0.4\pm { 0.8}$ \\ 
\#2  & $0.27\pm {0.13}$  & $ 1.7\pm { 3.0}$  & $ 1.0\pm { 1.6}$  & $ 3.7\pm { 2.9}$ \\ 
\hline 
\multicolumn{5}{c}{S\'ersic 159$-$3} \\ 
\#0  & $0.39\pm {0.03}$  & $ 0.6\pm { 0.4}$  & $ 1.3\pm { 0.2}$  & $ 0.8\pm { 0.3}$ \\ 
\#1  & $0.23\pm {0.03}$  & $ 0.8\pm { 0.8}$  & $ 1.4\pm { 0.6}$  & $ 0.5\pm { 0.5}$ \\ 
\#2  & $0.09\pm {0.05}$ & --  & $ 1.5\pm { 1.7}$  & $ 1.5\pm { 2.7}$ \\ 
\hline 
\multicolumn{5}{c}{MKW 9} \\ 
\#0  & $0.63\pm {0.15}$  & $ 0.1\pm { 0.5}$  & $ 1.9\pm { 0.7}$  & $ 1.4\pm { 0.8}$ \\ 
\#1  & $0.97\pm {0.45}$ & --  & $ 0.2\pm { 0.8}$  & $ 0.0\pm { 0.7}$ \\ 
\hline 
\multicolumn{5}{c}{2A 0335+096} \\ 
\#0  & $0.60\pm {0.04}$  & $ 1.6\pm { 0.7}$  & $ 1.0\pm { 0.2}$  & $ 0.8\pm { 0.3}$ \\ 
\#1  & $0.42\pm {0.07}$  & $ 1.8\pm { 1.7}$  & $ 1.2\pm { 0.6}$  & $ 0.0\pm { 0.4}$ \\ 
\#2  & $0.31\pm {0.17}$ & --  & $ 0.3\pm { 1.1}$  & $ 1.5\pm { 1.9}$ \\ 
\hline 
\multicolumn{5}{c}{MKW 3s} \\ 
\#0  & $0.68\pm {0.06}$  & $ 0.5\pm { 0.4}$  & $ 1.4\pm { 0.3}$  & $ 0.9\pm { 0.4}$ \\ 
\#1  & $0.31\pm {0.03}$  & $ 1.3\pm { 0.9}$  & $ 1.5\pm { 0.6}$  & $ 0.5\pm { 0.5}$ \\ 
\#2  & $0.09\pm {0.06}$  & $ 0.0\pm { 1.9}$ & -- & -- \\ 
\hline 
\multicolumn{5}{c}{A 2052} \\ 
\#0  & $0.58\pm {0.04}$  & $ 0.7\pm { 0.4}$  & $ 1.3\pm { 0.3}$  & $ 0.8\pm { 0.3}$ \\ 
\#1  & $0.41\pm {0.04}$  & $ 0.6\pm { 0.5}$  & $ 1.4\pm { 0.5}$  & $ 0.6\pm { 0.5}$ \\ 
\#2  & $0.13\pm {0.06}$ & --  & $ 0.3\pm { 1.1}$  & $ 3.3\pm { 2.6}$ \\ 
\hline 
\multicolumn{5}{c}{A 4059} \\ 
\#0  & $0.72\pm {0.12}$  & $ 0.5\pm { 0.6}$  & $ 1.3\pm { 0.6}$  & $ 1.7\pm { 0.8}$ \\ 
\#1  & $0.44\pm {0.08}$  & $ 0.7\pm { 0.7}$  & $ 1.4\pm { 0.9}$  & $ 0.7\pm { 0.8}$ \\ 
\#2  & $0.41\pm {0.23}$  & $ 1.2\pm { 2.5}$  & $ 3.2\pm { 2.2}$  & $ 1.7\pm { 2.5}$ \\ 
\hline 
\multicolumn{5}{c}{Hyd A} \\ 
\#0  & $0.42\pm {0.05}$  & $ 0.8\pm { 0.6}$  & $ 1.1\pm { 0.5}$  & $ 0.4\pm { 0.5}$ \\ 
\#1  & $0.30\pm {0.05}$  & $ 0.9\pm { 0.8}$  & $ 1.9\pm { 0.9}$  & $ 0.7\pm { 0.8}$ \\ 
\#2  & $0.20\pm {0.11}$ & --  & $ 0.5\pm { 1.7}$  & $ 0.0\pm { 1.2}$ \\ 
\hline 
\end{tabular}
\begin{list}{}{}
\item[$^{\mathrm{a}}$] Integration region (\#0, \#1, \#2 for radii
between 0--50, 50--200,
 and 200--500 in $h^{-1}$~kpc). 
\end{list}
\end{center}
\end{table}

\begin{table}[!ht]
\begin{center}
\caption{Abundances obtained from the EPIC analysis 
for each cluster (continued).
Same notation as Table~\ref{tbl:eack-abun}.}
\label{tbl:eack-abun2}
\begin{tabular}{rrrrr}
\hline 
R	& Fe/H	& O/Fe & Si/Fe & S/Fe\\
\hline 
\multicolumn{5}{c}{A 496} \\ 
\#0  & $0.60\pm {0.05}$  & $ 0.4\pm { 0.3}$  & $ 1.6\pm { 0.3}$  & $ 1.1\pm { 0.4}$ \\ 
\#1  & $0.37\pm {0.07}$  & $ 1.5\pm { 1.2}$  & $ 1.5\pm { 0.8}$  & $ 0.7\pm { 0.8}$ \\ 
\#2  & $0.41\pm {0.26}$ & --  & $ 1.7\pm { 2.3}$  & $ 1.8\pm { 2.7}$ \\ 
\hline 
\multicolumn{5}{c}{A 3112} \\ 
\#0  & $0.70\pm {0.05}$  & $ 0.9\pm { 0.6}$  & $ 1.4\pm { 0.4}$  & $ 0.4\pm { 0.3}$ \\ 
\#1  & $0.36\pm {0.05}$  & $ 0.2\pm { 0.7}$  & $ 1.0\pm { 0.8}$  & $ 0.6\pm { 0.7}$ \\ 
\#2  & $0.22\pm {0.13}$  & $ 0.0\pm { 1.6}$  & $ 1.3\pm { 2.3}$  & $ 2.3\pm { 2.7}$ \\ 
\hline 
\multicolumn{5}{c}{A 1795} \\ 
\#0  & $0.51\pm {0.04}$  & $ 0.7\pm { 0.4}$  & $ 1.4\pm { 0.4}$  & $ 0.6\pm { 0.4}$ \\ 
\#1  & $0.32\pm {0.04}$  & $ 0.5\pm { 0.7}$  & $ 1.9\pm { 1.1}$  & $ 0.0\pm { 0.5}$ \\ 
\#2  & $0.26\pm {0.07}$  & $ 1.4\pm { 1.9}$  & $ 3.2\pm { 2.1}$  & $ 0.6\pm { 1.2}$ \\ 
\hline 
\multicolumn{5}{c}{A 399} \\ 
\#0  & $0.63\pm {0.35}$  & $ 0.0\pm { 2.2}$  & $ 0.0\pm { 1.2}$  & $ 0.0\pm { 1.4}$ \\ 
\#1  & $0.31\pm {0.12}$ & --  & $ 0.9\pm { 2.2}$ & -- \\ 
\#2  & $0.28\pm {0.13}$ & -- & -- & -- \\ 
\hline 
\multicolumn{5}{c}{A 3266} \\ 
\#0  & $0.46\pm {0.34}$ & -- & -- & -- \\ 
\#1  & $0.27\pm {0.13}$ & -- & -- & -- \\ 
\#2  & $0.22\pm {0.08}$ & -- & -- & -- \\ 
\hline 
\multicolumn{5}{c}{Perseus} \\ 
\#0  & $0.57\pm {0.03}$  & $ 1.4\pm { 0.3}$  & $ 1.2\pm { 0.2}$  & $ 0.4\pm { 0.4}$ \\ 
\#1  & $0.41\pm {0.04}$  & $ 2.3\pm { 0.9}$  & $ 1.2\pm { 0.4}$  & $ 0.2\pm { 0.3}$ \\ 
\hline 
\multicolumn{5}{c}{Coma} \\ 
\#0  & $0.53\pm {0.54}$ & -- & -- & -- \\ 
\#1  & $0.26\pm {0.09}$ & --  & $ 3.8\pm { 2.8}$  & $ 0.7\pm { 2.0}$ \\ 
\#2  & $0.25\pm {0.10}$  & $ 2.4\pm { 2.8}$  & $ 1.4\pm { 2.1}$  & $ 0.0\pm { 0.9}$ \\ 
\hline 
\multicolumn{5}{c}{A 754} \\ 
\#0  & $0.32\pm {0.34}$  & $ 0.0\pm { 2.5}$ & -- & -- \\ 
\#1  & $0.33\pm {0.11}$  & $ 1.1\pm { 2.2}$  & $ 0.4\pm { 1.4}$  & $ 0.6\pm { 1.6}$ \\ 
\#2  & $0.21\pm {0.09}$ & -- & -- & -- \\ 
\hline 
\multicolumn{5}{c}{A 1835} \\ 
\#0  & $0.42\pm {0.06}$  & $ 2.0\pm { 1.4}$  & $ 1.1\pm { 0.9}$  & $ 0.6\pm { 1.0}$ \\ 
\#1  & $0.38\pm {0.09}$  & $ 0.0\pm { 1.0}$  & $ 0.0\pm { 0.6}$  & $ 0.0\pm { 1.1}$ \\ 
\#2  & $0.29\pm {0.11}$ & -- & --  & $ 1.7\pm { 2.8}$ \\ 
\hline
\end{tabular}
\end{center}
\end{table}


\begin{table*}[!ht]
\begin{center}
\caption{Average abundances obtained from the EPIC analysis for the cool, 
medium-temperature, hot, and soft excess clusters.
Same notation as Table~\ref{tbl:eack-abun}.
}
\label{tbl:epic-abun-av}
\begin{tabular}{rrrrrrrrr}
\hline
R	& Fe/H	& O/Fe & Ne/Fe & Mg/Fe & Si/Fe & S/Fe & Ar/Fe & Ca/Fe \\
\hline 
\multicolumn{9}{c}{Cool ($kT< 3$~keV)} \\
\#0  & $0.57\pm {0.04}$  & $ 0.4\pm { 0.2}$  & $ 1.4\pm { 0.4}$  & $ 0.6\pm { 0.3}$  & $ 1.6\pm { 0.2}$  & $ 1.4\pm { 0.3}$  & $ 1.9\pm { 1.5}$  & $ 3.1\pm { 2.1}$ \\ 
\#1  & $0.42\pm {0.08}$  & $ 1.9\pm { 0.9}$  & $ 1.0\pm { 0.8}$  & --                & $ 1.2\pm { 0.5}$  & $ 0.7\pm { 0.6}$ & -- & -- \\ 
\hline
\multicolumn{9}{c}{medium-T ($3<kT< 6$~keV)} \\
\#0  & $0.60\pm {0.04}$  & $ 0.7\pm { 0.4}$  & $ 1.8\pm { 0.5}$  & $ 0.3\pm { 0.3}$  & $ 1.6\pm { 0.3}$  & $ 1.0\pm { 0.3}$  & $ 1.1\pm { 0.7}$  & $ 2.9\pm { 1.3}$ \\ 
\#1  & $0.39\pm {0.03}$  & $ 1.2\pm { 0.5}$  & $ 1.3\pm { 0.5}$  & --                & $ 1.4\pm { 0.4}$  & $ 0.5\pm { 0.3}$  & $ 1.2\pm { 1.0}$  & $ 3.3\pm { 1.6}$ \\ 
\#2  & $0.32\pm {0.08}$  & $ 4.0\pm { 2.5}$  & $ 1.5\pm { 1.2}$  & --                & $ 1.3\pm { 0.9}$  & $ 1.7\pm { 1.1}$  & $ 1.3\pm { 2.8}$ & -- \\ 
\hline
\multicolumn{9}{c}{Hot ($kT> 6$~keV)} \\
\#0  & $0.48\pm {0.12}$  & $ 1.2\pm { 1.2}$  & $ 2.6\pm { 1.9}$  & $ 1.3\pm { 2.2}$  & $ 2.2\pm { 1.9}$  & $ 0.5\pm { 1.6}$ & -- & -- \\ 
\#1  & $0.34\pm {0.05}$  & $ 2.1\pm { 1.7}$  & $ 3.6\pm { 2.0}$  & --                & $ 1.9\pm { 1.7}$  & $ 1.8\pm { 1.8}$  & $ 3.3\pm { 4.1}$ & -- \\ 
\#2  & $0.25\pm {0.05}$ & --  & $ 3.3\pm { 3.4}$  & --  &  --          & $ 3.1\pm { 4.0}$ & -- & -- \\ 
\hline
\multicolumn{9}{c}{Soft excess clusters} \\
\#0  & $0.57\pm {0.02}$  & $ 0.6\pm { 0.2}$  & $ 1.6\pm { 0.3}$  & $ 0.4\pm { 0.2}$  & $ 1.4\pm { 0.2}$  & $ 0.7\pm { 0.2}$  & $ 0.8\pm { 0.4}$  & $ 2.4\pm { 0.7}$ \\ 
\#1  & $0.32\pm {0.02}$  & $ 0.8\pm { 0.3}$  & $ 0.8\pm { 0.4}$  & --                & $ 1.5\pm { 0.3}$  & $ 0.4\pm { 0.3}$  & $ 1.1\pm { 0.8}$  & $ 3.1\pm { 1.3}$ \\ 
\#2  & $0.17\pm {0.04}$  & $ 1.5\pm { 1.2}$  & $ 2.6\pm { 1.2}$  & --                & $ 2.2\pm { 1.0}$  & $ 1.8\pm { 1.1}$  & $ 1.1\pm { 3.3}$ & -- \\ 
\hline\\
\end{tabular}
\end{center}
\end{table*}

\begin{figure}[!ht]
\begin{center}
	\resizebox{\hsize}{!}{\includegraphics[angle=-90]{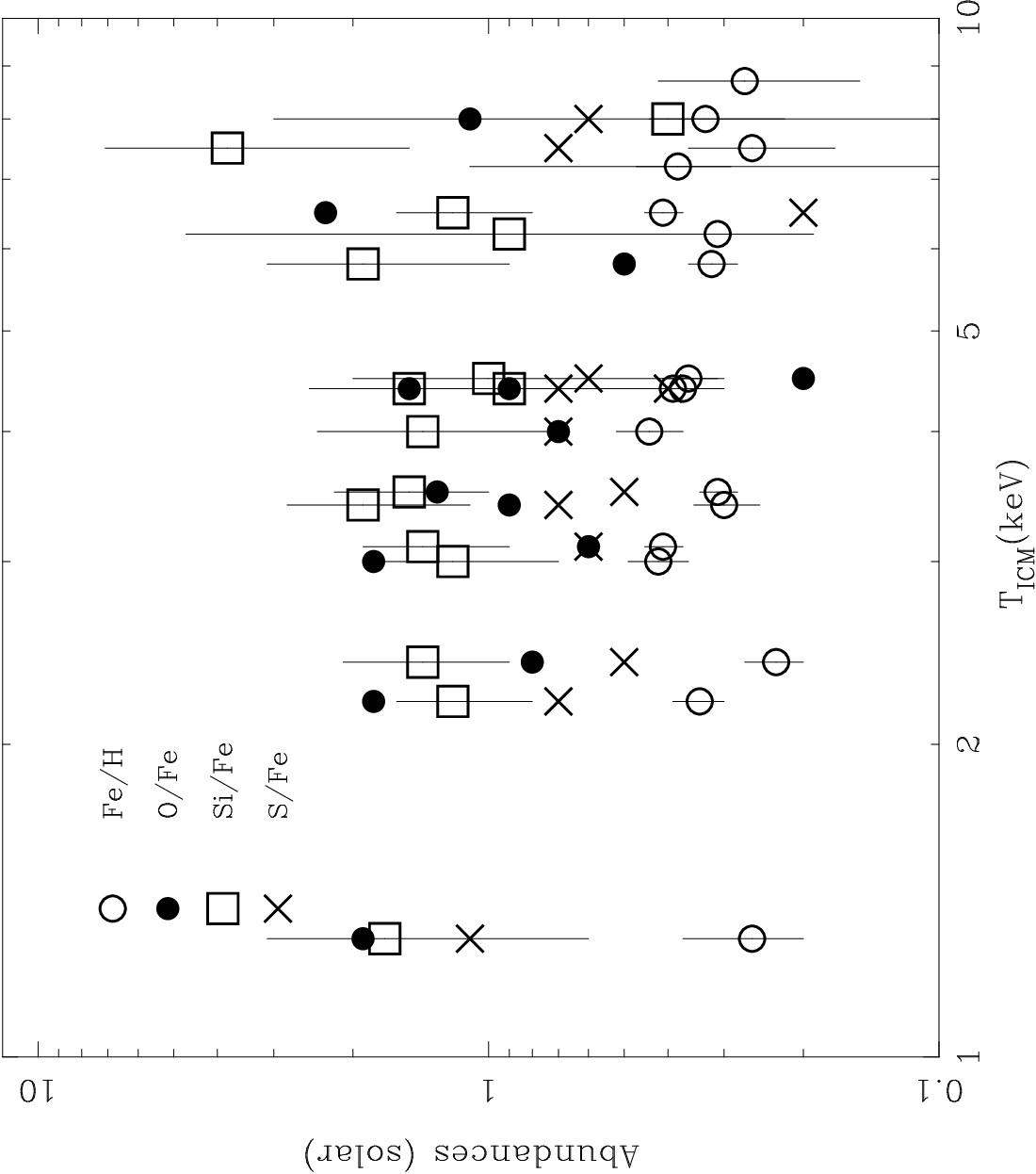}}
\caption{Abundance vs temperature for each cluster averaged 
over (50--200)~$h^{-1}$~kpc. Errors on O/Fe and S/Fe are not shown
for clarity (These are presented in 
Tables~\ref{tbl:eack-abun}--\ref{tbl:eack-abun2}). 
The data for MKW~9 are not shown for clarity, 
because this cluster has larger errors than the other clusters.
}
\label{fig:tem-metal}
\end{center}
\end{figure}

In all cases except for A~3226, Coma and A~754, the temperature decreases
towards the cluster center (Kaastra et al.  \cite{kaastra04}).  In
Fig.~\ref{fig:epic-radial-ab}, we present the abundances as derived from the
acceptable fits ($\chi^2/\nu <1.3$) of either the 1T or 2T models.  In
Tables~\ref{tbl:eack-abun}--\ref{tbl:eack-abun2}, we show average values over
three different spatial scales.  We also present 
average values for cool ($kT_{\rm ICM} < 3$~keV), medium-temperature, and hot
clusters ($kT_{\rm ICM} > 6$~keV) in Table~\ref{tbl:epic-abun-av}.  Furthermore, we separately show the results
for some clusters which exhibit soft excess (''SE'' in Table~\ref{tbl:sample}).
Fig.~\ref{fig:tem-metal} shows abundances averaged over (50--200)$h^{-1}$~kpc
for each cluster as a function of the ICM temperature.  Below we summarize the
main results of the Fe abundance, O/Fe, Si/Fe, and S/Fe ratios.


In all cool and medium-temperature clusters except
MKW 9, the faintest one,
we detected a central increase in the Fe abundance; the abundance
starts to increase at (50--100)~$h^{-1}$~kpc in radius from 0.2--0.4~solar to
0.6--0.8~solar towards the center.  
Except for the Perseus cluster, we found no significant Fe increase in the inner regions of the hot clusters (See Table \ref{tbl:eack-abun2}). 
However, the best-fit value for each cluster suggests the Fe increase in all cases and the average Fe value over the hot clusters also indicates a central enhancement (Table~\ref{tbl:epic-abun-av}.)

In the outer regions, (200-500)~$h^{-1}$~kpc
in radius for all clusters, the Fe abundance is 0.2--0.3, without a significant
dependence on the ICM temperature (Fig.~\ref{fig:tem-metal}).  Exceptions are
the soft excess clusters, which show relatively smaller Fe abundances down to
0.1~solar.  These low Fe abundances may not be the actual values due to a spectral
contamination by the soft excess emission, 
which was not taken into account in the present analysis.
Further careful analysis is necessary to take into account the soft excess emission and derive the accurate Fe abundance. This is beyond the scope of the present paper.


The error on the Si abundance determination is larger than on iron.
Nevertheless, there is also an indication of a central increase in the Si
abundance.  In fact, the Si/Fe ratio is radially uniform at an average value of
1--2 solar (Fig.~\ref{fig:epic-radial-ab}c).  Within the error bars, there is no
radial change of the Si/Fe ratio for each cluster and among different clusters
(Fig.~\ref{fig:tem-metal}).  The radial distribution of the S abundance is
similar to that of Si; the S/Si ratios are consistent with the solar ratio.


In the cool and medium-temperature clusters, the measured O/Fe abundance ratio
at the cluster center is 0.4--0.7~relative to the solar value.  This is consistent with the values
obtained from the RGS (as shown later).  For the hot clusters, the errors on the
O abundance are too large to allow a measurement of spatial variations.
Contrary to Fe and Si, the O abundance shows no significant central increase.
The O/Fe ratio averaged over the cool and medium-temperature clusters increases
towards large radii to a value of a few times solar.


The Ne, Mg, Ar, and Ca abundances have larger errors.  We only show the average
abundance of these elements in Table~\ref{tbl:epic-abun-av}.  In general,
these abundances relative to Fe are consistent with the solar ratio.

\section{The RGS analysis}

\subsection{Analysis method}

The RGS events were filtered using both the dispersion vs.  cross-dispersion and
dispersion vs.  pulse height windows.  We used a spectral extraction region of
1\arcmin.0 in full-width in the cross-dispersion direction centered on the X-ray
maximum.  This corresponds to a physical size of 50~$h^{-1}$~kpc for a mean
redshift of 0.06.  
Generally, most of the source photons in the RGS aperture are
inside this extraction region.  
We used a pulse height width selection that included
90\% of all photons for a point source.

Our targets have a larger spatial extent than the instrumental spatial
resolution, resulting in a broadening of the line spread function (LSF).  Here
we modeled the source brightness using the observed MOS image for each target.
Then, we corrected for this effect by convolving the cluster brightness profile
with the RGS response for a point source.  Because the dependence of the LSF on
the off-axis angle within $\pm3$\arcmin\ is smaller than 10--30\% and the
cluster emission is dominated by emission within that spatial range, we ignore
this dependence of the LSF.

We have estimated  the background spectrum and its variability using several
blank-sky field observations ($\sim 400$~ks in total exposure; Tamura et al. \cite{tamura03}).
The count rate variations
between the different exposures
is small compared to the statistical errors:
the r.m.s. deviations over the full spectral range are less than 30\%.
Based on this, a 30~\% systematic error to the background counts was assigned.

The first order spectra from the RGS1 and RGS2 were fitted with the same model
simultaneously.  We limited the wavelength band to 8--28~\AA, where the
estimated background is typically less than 10~\% of the source counts.
Therefore, the systematic uncertainty in the background modeling should not
affect the results significantly.

In principle, the extracted spectrum contains photons from a range of different
sky positions.  However, in the cases of our sample, where the brightness is
centrally peaked, the observed spectrum is dominated by the photons from the
cluster core.  By using a full instrument simulation, SciSim (the official
{\it XMM-Newton} science simulation tool\footnote{see http://xmm.vilspa.esa.es}) and
the observed MOS image, we confirmed that the contamination from the cluster
outer region (a projected radius from the cluster center larger than 3\arcmin)
is less than 10~\% for a typical cluster and 20~\% at most for all clusters.
This kind of contamination adds counts in the form 
of a continuum spectrum since lines are smeared out due to
the large spatial extent of the outer cluster regions.
This could make the obtained equivalent line width and hence the metal abundance relative
to H appeared lower than the actual value by less than 10--20~\%.  The
possible change of abundances relative to Fe (e.g.  O/Fe) should be much
smaller.  Since the thermal structure is mainly determined by the Fe-L line
ratios, the effect of the contamination on the thermal structure should also be
small.  Therefore, we ignored this flux mixing and assumed that the values
obtained from our spectral fits are emission-weighted values for the cluster
core.

\subsection{Single-temperature model fitting}

We fitted a set of spectra for each cluster with the single temperature (1T)
model.  With this case, the free parameters are the normalization, $kT$, the Fe
abundance, and the Fe/O ratio.  An exception is NGC~533, where the RGS spectrum
is dominated by line emission.  In this case, it is relatively difficult to
determine the continuum level and hence the line equivalent widths.  Therefore
for NGC~533 we fixed the Fe abundance to the value obtained from the EPIC
analysis above.

We obtained
acceptable or almost acceptable fits ($\chi^2/\nu < 1.2$) with the 1T model for
S\'ersic~159$-$3, A~4059, Hyd-A, A~496, A~3112, A~1795 and A~1835.  In
contrast, NGC~533 ($\chi^2/\nu = 1.55$), A~262 (1.49), A~2052 (1.32), and 2A~0335 (1.48) exhibit clear deviations from the 1T model.

\subsection{Multi-temperature model fitting}


In some clusters, there is evidence for multi-temperature emission (Peterson et
al.~\cite{peterson03}).  If there is a significant contribution from the cooler
temperature components in the line emission, the abundances derived from the 1T
model may differ from the true abundances. 


To avoid this error, we used a multi-temperature model which consists of at most
five temperature CIE components with fixed temperature separation, ($T_0 = 2T_1
= 4T_2 = 8T_3 = 16T_4$), but free normalization ($EM_i$) for each component (the 5T model); 
a similar approach was also taken by Peterson et al. (\cite{peterson03}).
In this case, the Ne/Fe and Mg/Fe ratios are also allowed to be free along with
the Fe/H and O/Fe ratios.  All temperature components were assumed to have the
same abundances.


\begin{table*}[!ht]
\caption[]{RGS fit results for a multi-temperature model.
The mean abundances for the sample are also shown in the last row.}
\label{tbl:rgs-fits-5t}
\begin{center}
\begin{tabular}{lcccccc}
\hline 
target	& $kT_0^{\mathrm{a}}$ & O/Fe & Ne/Fe & Mg/Fe & Fe/H	
	& $\chi^2/\nu$ \\
\hline
NGC~533         & $ 1.4\pm 0.5$& $ 0.54\pm0.15$& $  1.3\pm 0.7$& $  1.7\pm 1.0$& $ 0.50       $& 264/226 \\
A~262           & $ 2.4\pm 0.7$& $ 0.64\pm0.12$& $  0.6\pm 0.5$& $  0.1\pm 0.5$& $ 0.58\pm0.11$& 229/231 \\
S\'ersic~159$-$3& $ 2.3\pm 0.2$& $ 0.67\pm0.15$& $  0.8\pm 0.4$& $  0.6\pm 0.5$& $ 0.38\pm0.03$& 279/263 \\
A~2052          & $ 2.8\pm 0.3$& $ 0.66\pm0.10$& $  0.8\pm 0.3$& $  0.9\pm 0.6$& $ 0.68\pm0.08$& 284/263 \\
2A~0335         & $ 1.6\pm 0.1$& $ 1.00\pm0.23$& $  0.4\pm 0.5$& $  1.0\pm 0.6$& $ 0.29\pm0.04$& 282/188 \\
A~4059          & $ 3.5\pm 0.7$& $ 0.68\pm0.16$& $  1.4\pm 0.7$& $  0.6\pm 0.8$& $ 0.58\pm0.10$& 215/181 \\
Hydra A         & $ 2.9\pm 0.3$& $ 0.70\pm0.18$& $  0.5\pm 0.6$& $  0.0\pm 0.3$& $ 0.37\pm0.06$& 230/228 \\
A~496           & $ 3.2\pm 0.6$& $ 0.66\pm0.15$& $  1.2\pm 0.6$& $  1.1\pm 0.9$& $ 0.55\pm0.14$& 236/229 \\
A~3112          & $ 4.2\pm 0.8$& $ 0.63\pm0.18$& $  0.5\pm 0.6$& $  0.0\pm 0.4$& $ 0.69\pm0.14$& 189/225 \\
A~1795          & $ 6.4\pm 1.0$& $ 0.78\pm0.14$& $  1.5\pm 0.4$& $  1.4\pm 0.8$& $ 0.60\pm0.13$& 284/261 \\
A~1835          & $ 4.6\pm 1.0$& $ 0.35\pm0.24$& $  0.9\pm 0.6$& $  0.0\pm 0.7$& $ 0.49\pm0.12$& 295/284 \\
\hline
mean	        & -	       & $ 0.63\pm0.05$&  $ 1.0\pm 0.2$& $ 0.66\pm 0.2$& $ 0.54\pm0.03$& -- \\ 
\hline
\end{tabular}
\begin{list}{}{}
\item[$^{\mathrm{a}}$] Temperature of the hottest component in keV.
\end{list}
\end{center}
\end{table*}

The results are shown in Table~\ref{tbl:rgs-fits-5t}.  In S\'ersic~159$-$3,
A~4059, Hyd-A, A~3112 and A~1835 where the 1T model already provided an acceptable fit,
this model does not improve the fit.  The derived abundances are consistent with
those of the 1T fit, simply because the same temperature component dominates the
emission in both the 1T and 5T models.

In 2A~0335+096, we could not obtain an acceptable fit even with the 5T model.
We note that the EPIC spectra of the inner region of this cluster also show a
deviation from the 2T model (Table~\ref{tbl:in-fit}).  The Galactic absorption
towards this cluster is by far larger than that of other clusters.  Furthermore, the
X-ray image shows a complex structure (Sarazin et al.  \cite{sarazin92}).
Therefore, we assumed that our spectral and spatial modeling is oversimplified
for this cluster.  Note that we exclude the EPIC results for the inner region of
this cluster in our discussion.

In other clusters, the 5T model improves the fit significantly.  
The best example is A~262, where the model reproduced the
observed spectra very well.  By using this model, we obtained acceptable fits
($\chi^2/\nu < 1.2$) for all clusters.  While there are significant changes in
the derived Fe abundance in some cases, the change in the O/Fe ratio is at most
30~\% or 0.2 times solar, compared to the 1T fitting.

Peterson et al.  (\cite{peterson03}) analyzed a similar RGS data set, focusing
on the temperature structure.  Our model for the temperature structure is
basically consistent with their results.

In Fig.~\ref{fig:epic_vs_rgs}, we compare the O and Fe abundance as obtained by
EPIC and RGS.  
In 2A~0335+096, the Fe/H ratio derived from the RGS data is two times
smaller than that derived from the EPIC data, while the O/Fe ratio from the two
instruments is consistent.  We presume that this discrepancy is due to the
nature of the central spectra which is more complex than our modeling.
Although the errors in the oxygen abundance determination from EPIC is large, 
apart from 2A~0335+096, the two instruments provide almost consistent
results: 
a formal fit of the Fe abundance from RGS on the Fe abundance from EPIC to the 1:1 relation yields $\chi^2/\nu=7.68/10$.

\section{Summary and Discussion}

\subsection{Present results}

Using spatially-resolved EPIC spectra, we have resolved the temperature and
abundance distribution on a sub-arcmin scale in a sample of 19 clusters.
Furthermore, using high spectral resolution RGS spectra of 11 cluster cores, we have
resolved the soft X-ray spectrum dominated by the Fe-L and the
\ion{O}{viii}~Ly$\alpha$ lines.  This provides not only tight constrains on the
thermal structure but also a robust measurement of the O abundance.  
Below we summarize our main results.

\begin{enumerate}
 
\item
We found no significant variation in the Fe abundance and the Si, S, and O ratios to the Fe among the systems (see Fig.~\ref{fig:tem-metal}).

\item 
The O/H and O/Fe ratio in the cluster cores, (20--60)~$h^{-1}$~kpc in
radius, are 0.34$\pm 0.03$ and 0.63$\pm 0.05$, respectively.  The r.m.s.
deviation for individual clusters as compared to these average values is
comparable to the measurement errors on these ratios.

\item 
In all clusters with a temperature less than 6~keV, except for the faint
cluster MKW~9, we detected a central increase in the Fe abundance; the abundance
starts to increase at (50--100)~$h^{-1}$~kpc in radius from 0.2--0.4~solar value to
0.6--0.8~solar value towards the center.

\item 
The Si/Fe and S/Fe ratios in cool and medium temperature clusters are
radially uniform within (200--500)~$h^{-1}$~kpc with mean values of 1.4$\pm$0.2
and 1.1$\pm$0.3, respectively, relative to the solar ratio.

\item 
Contrary to the Fe, Si, and S abundances, the O abundance shows no spatial
variations. 
The errors on the O abundance for each cluster are large.
However, when we combine the results from several clusters and assume no correlation of the systematic errors of the measurements between clusters, 
we detected a significant radial variation in the O/Fe ratio; 
the mean O/Fe ratios are $0.5^{+0.3}_{-0.1}$, $1.7^{+0.9}_{-0.4}$,
and $4.1^{+3.5}_{-1.9}$ for (0--50)~$h^{-1}$~kpc,(50--200)~$h^{-1}$~kpc, and
(200--500)~$h^{-1}$~kpc, respectively.

\end{enumerate}

\subsection{Similarity of the abundance among clusters}
{\it ASCA} and {\it BeppoSAX} observations have established the Fe increase towards the cluster center 
in relatively cool clusters with cD galaxies (e.g. Fukazawa et al. \cite{fukazawa00}; De Grandi and Molendi \cite{grandi01}).
Our measurement not only confirms these results in the cool and medium temperature clusters but also suggests a similar enhancement in hot clusters without a prominent cD galaxy (e.g. Coma), probably because of the improved spatial resolution.

Beside the Fe abundance, we found no variation in the Si and other element ratios to Fe among the clusters. This implies a homogeneous production of the metals in our sample.
However, the sample is biased towards medium-temperature X-ray luminous clusters.
Therefore we make no conclusion about the universal distribution of the relative abundances.
In fact Fukazawa et al. (\cite{fukazawa98}) claimed an increase in Si/Fe from cool to hot systems using 40 clusters.

\subsection{The O abundance at the cluster core}

The theoretical models for supernova (SN) types Ia and II predict the O/Fe ratio
to be $<0.05$ and $1.5-4$ times solar values, respectively.  On the other hand,
the observed O/Fe ratio at the cluster cores are between these two predicted
values.  This suggests that the ICM metals at the cluster cores could be produced by a
mix of both types of SNe.  
A value of the O/Fe abundance ratio in the ICM close
to solar values supports the idea that the metals in the ICM and the Galaxy were produced
in a similar way. 
Assuming the O and Fe yields of SN Ia and II from Tsujimoto et al.  (\cite{tsujimoto95}), the observed O/Fe ratio ($\sim 0.65$) can be reproduced by the relative frequencies of the two SNe, $N_{\rm Ia}/N_{\rm II}$, of $\sim 0.6$. 
In this model, the mass fractions of the O and Fe originated from SN Ia are 0.05 and $\sim 0.8$, respectively.

The observed O abundance provides an accurate way to determine the total mass of
O in the ICM, $M_{\rm O}^{\rm ICM}$.  We estimate $M_{\rm O}^{\rm ICM}$ within
50~$h^{-1}$ kpc to range from $10^8$ (NGC~533) to $2\times 10^9$ (A~1795) in
units of $h^{-2.5}$~M$_{\odot}$ for our RGS sample.  We assume that all oxygen
was produced in SNe II.  Then, combining $M_{\rm O}^{\rm ICM}$ with a SN II
model O yield ($\sim 2$~M$_{\odot}$, Tsujimoto et al.  \cite{tsujimoto95}),
gives a total number of SNe II of $10^8-2.5\times10^9$ for $h=0.7$.  Further
assuming that the duration of SNe II activity is $10^7$~yr ( a typical life time
of a 20~M$_{\odot}$ star), we can estimate the SN II rate to be 10--200~SN
IIe~yr$^{-1}$.  This rate is at least one or two orders of magnitude larger than
that of a typical starburst galaxy.  In other words, the observed O mass
requires a large amount of massive star formation, producing
the corresponding number of SNe II.

The observed O/H ratio in cluster cores (20--60~kpc in radius) is about half
that observed in the central regions (5--15~kpc) of giant elliptical galaxies,
like in M87 (Sakelliou et al.  \cite{sakelliou02}; Gastaldello and Molendi
\cite{gas02}; Finoguenov et al.  \cite{finoguenov02}) and NGC~4636 (Xu et al.
\cite{xu02}).  The overabundance of oxygen in galaxy centers suggests that in
those regions oxygen has been produced via massive star formation and that
metals have not been completely mixed with the ambient gas (Gastaldello and
Molendi \cite{gas02}).

\subsection{The total Fe mass in clusters}

\begin{figure}[!ht]
\begin{center}
\resizebox{\hsize}{!}{\includegraphics[angle=-90]{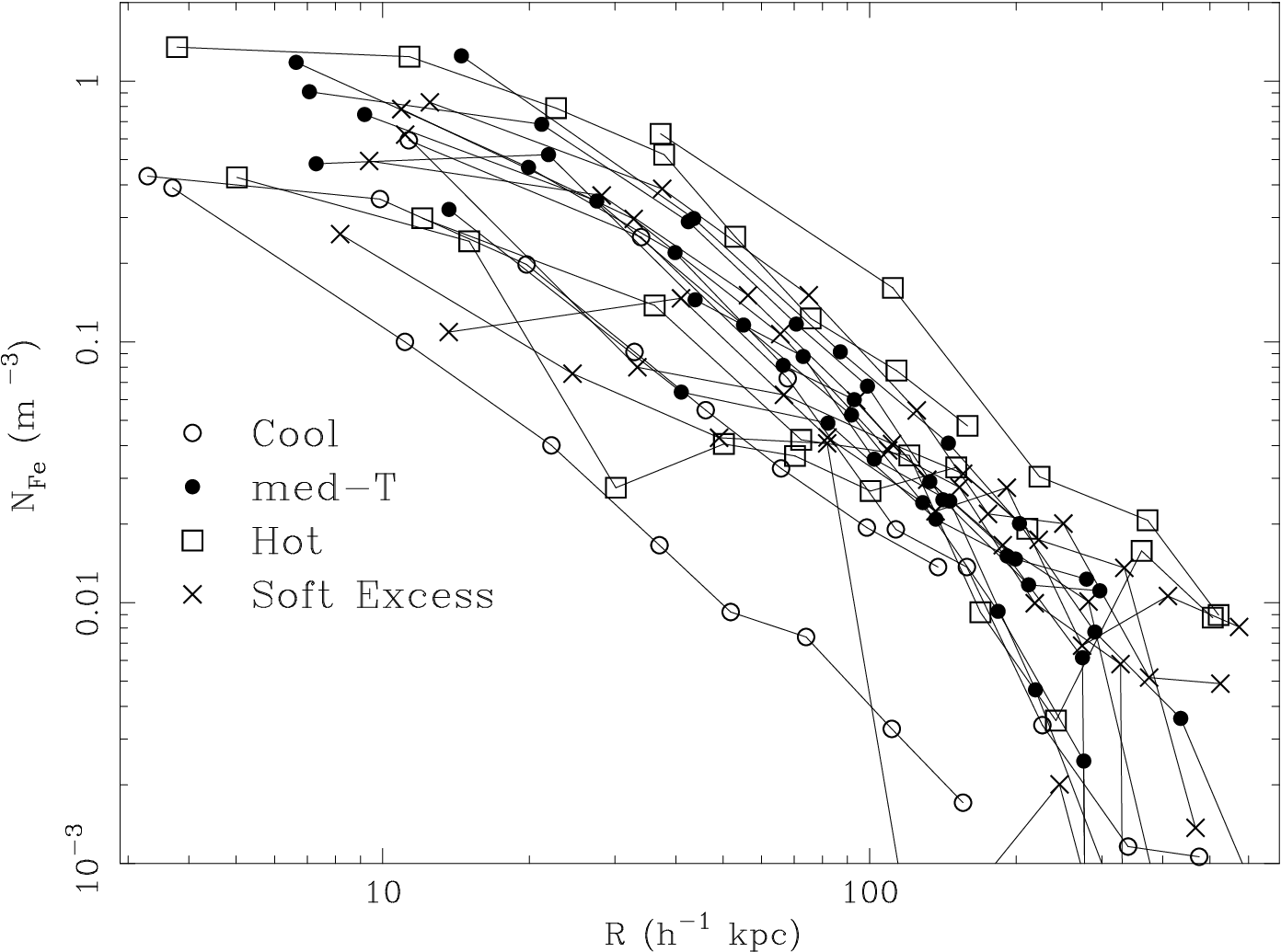}}
\caption{The Fe number density in the ICM. 
Same notation as Fig.~\ref{fig:epic-radial-ab}.}
\label{fig:femass1}
\resizebox{\hsize}{!}{\includegraphics[angle=-90]{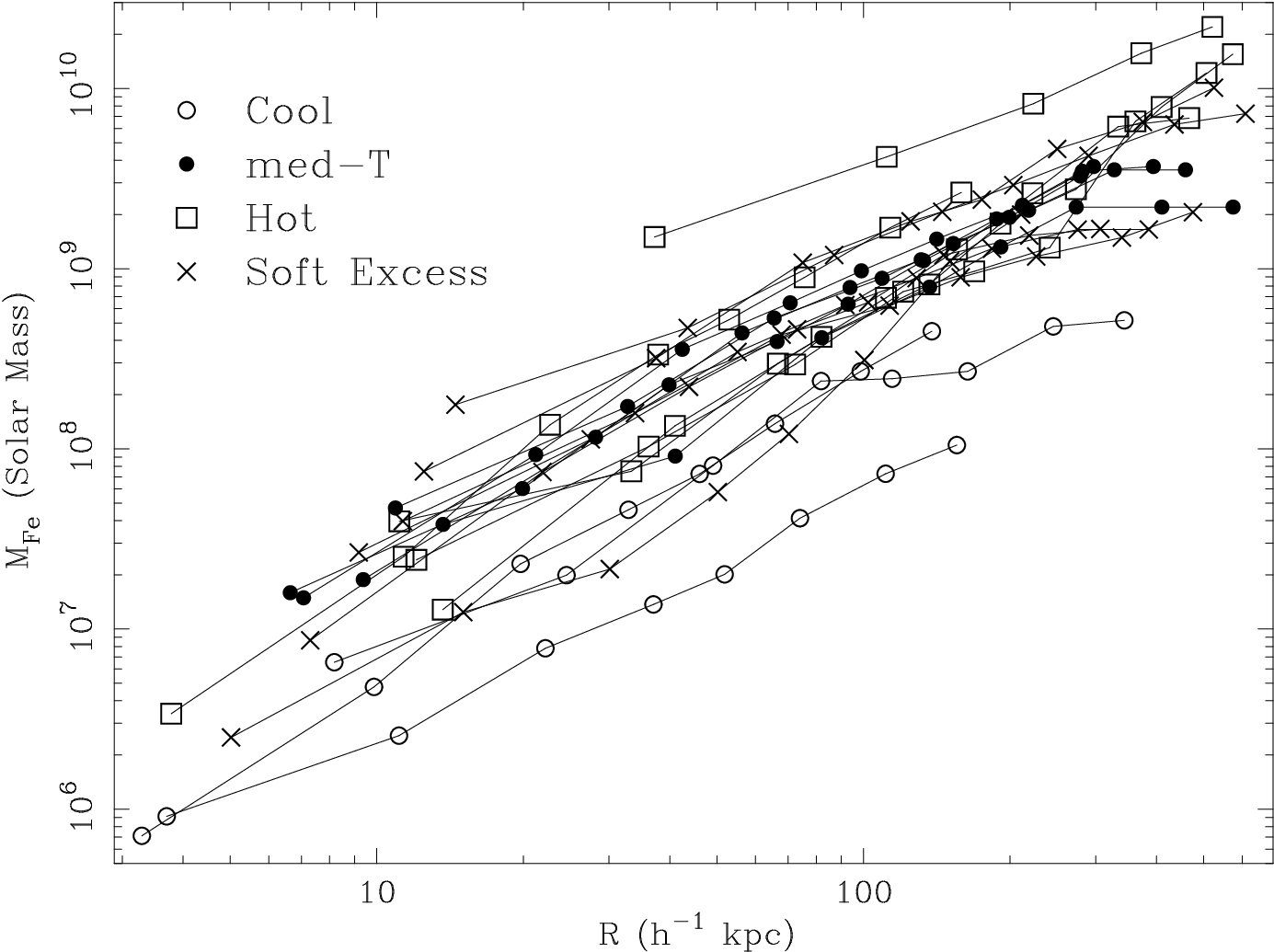}}
\caption{Same as the previous figure, but for the integrated Fe mass in the ICM. }
\label{fig:femass2}
\end{center}
\end{figure}

\begin{table}[!ht]
\caption{The iron mass to light ratio within $250~h^{-1}$~kpc.}
\label{tbl:imlr}
\begin{center}
\begin{tabular}{lccc}
\hline 
cluster & $M_{\rm Fe} $ & $N_{\rm 0.5}^{\mathrm{a}}$ & $M_{\rm Fe}/L_{\rm B}$ \\
	& ($10^9$ M$_{\odot}$) & & ($10^{-2}$M$_{\odot}$/L$_{\odot}$)\\
\hline
MKW~9	& 0.5  	& 13	& 0.3 $\pm 0.15$\\
MKW~3s	& 1.5	& 13	& 0.9 $\pm 0.2$\\
A~496	& 2.8	& 14	& 1.5 $\pm 0.5$\\
A~1795	& 6.0	& 27	& 1.0 $\pm 0.2$\\
A~754	& 3.8	& 29	& 0.7 $\pm 0.3$\\
\hline
\end{tabular}
\begin{list}{}{}
\item[$^{\mathrm{a}}$] Number of galaxies within a radius of 250~$h^{-1}$~kpc.
\end{list}
\end{center}
\end{table}

We also calculated the total Fe density and mass in the ICM as a function of
radius, based on our measurements of the emission measure and the Fe abundances
(Figs.~\ref{fig:femass1} and \ref{fig:femass2}).  To compare this Fe mass and
the predicted Fe production rate in galaxies, we estimate the iron mass to light
ratio (IMLR).  We use the galaxy number within a radius of 250~$h^{-1}$~kpc,
$N_{0.5}$ (Bahcall \cite{bahcall81} and reference therein), and estimate the
total optical luminosity of member galaxies based on the relation $L_B =
4.5\times 10^9 N_{0.5}^{1.42} h^{-2}$~L$_{\odot}$ (Edge and Stewart
\cite{edge91}).  Table~\ref{tbl:imlr} shows the results for representative
clusters, where $N_{0.5}$ is available.

The IMLR of MKW~9 is relatively smaller than for other clusters.  
The potential of MKW 9 might not be as deep as in other systems
and a large part of the metal-enriched gas has been driven out already from the potential
well.  An IMLR of order $10^{-2}$ in other systems is consistent with previous
measurements (e.g.  Tsuru \cite{tsuru91}) and at least a few times larger than
the predicted value based upon the current rate of SN Ia in elliptical galaxies,
as argued previously (e.g., Tsuru \cite{tsuru91}; Renzini et al.
\cite{renzini93}).  We presume that a large fraction of Fe has been produced
also in the early phase of the galaxy formation with a high production rate.
Relatively low metal abundances observed in elliptical galaxies (sub-solar at
most; e.g.  Matsushita et al.  \cite{matsushita00}) also support this idea.

\subsection{Spatial distribution of the O, Si, S, and Fe abundances}
The most simple model to explain the results (3) and (4) is that Si, S, and Fe have a common origin for a large part, while O has a different origin.  
This contrast between Si-S-Fe and O
has reported separately in A~496 (Tamura et al.  \cite{tamura01b}).
Qualitatively this trend is consistent with the current idea that the
contribution from SN Ia relative to that from SN II, $N_{\rm Ia}/N_{\rm II}$,
increases towards the cluster center and the O/Fe ratio is more sensitive to
$N_{\rm Ia}/N_{\rm II}$ than the Si/Fe ratio.  We do not attempt a further
detailed comparison between the observation and models, because of large
uncertainties in both the data (e.g.  the O/Fe ratio in the outer regions and
for hot clusters) and the theory (e.g.  Gibson et al.  \cite{gibson97}).

\begin{acknowledgements}
This work is based on observations obtained with {\it XMM-Newton}, an ESA science
mission with instruments and contributions directly funded by ESA Member States
and the USA (NASA).  The Space Research Organisation of the Netherlands (SRON)
is supported financially by NWO, the Netherlands Organisation for Scientific
Research.
We thank the referee for useful comments.
\end{acknowledgements}


\begin{thebibliography}{}
\bibitem[1989]{anders}
  Anders, E., Grevesse, N., 1989, Geochim. Cosmochim. Acta 53, 197
\bibitem[1992]{arnaud92} Arnaud, M., Rothenflug, 
R., Boulade, O., Vigroux, L., \& Vangioni-Flam, E.\ 1992, A\&A, 254, 49
\bibitem[1981]{bahcall81} 
  Bahcall, N.~A.\ 1981, ApJ, 247, 787  
\bibitem[2000]{buote00}
  Buote, D.A., 2000, MNRAS, 311, 176
\bibitem[2001]{david01} David, L.~P., Nulsen, 
P.~E.~J., McNamara, B.~R., Forman, W., Jones, C., Ponman, T., Robertson, 
B., \& Wise, M.\ 2001, ApJ, 557, 546
\bibitem[1990]{dickey}
Dickey, J.M., \& Lockman, F.J., 1990, ARA\&A 28, 215
\bibitem[1991]{edge91} 
  Edge, A.C., \& Stewart, G.C. 1991, MNRAS, 252, 428
\bibitem[2000]{finoguenov00} 
Finoguenov, A., David, L.P., Ponman, T.J., 2000, ApJ, 544, 188
\bibitem[2002]{finoguenov02} 
  Finoguenov, A., Matsushita, K., B{\" o}hringer, H., Ikebe, Y.,
  \& Arnaud, M.\ 2002, A\&A, 381, 21
\bibitem[1998]{fukazawa98} 
Fukazawa, Y., Makishima, K., Tamura, T., et al. 1998, PASJ, 50, 187
\bibitem[2000]{fukazawa00} 
Fukazawa, Y., Makishima, K., Tamura, T., Nakazawa, K., Ezawa, H., Ikebe, Y., Kikuchi, K., 
\& Ohashi, T.\ 2000, \mnras, 313, 21 
\bibitem[2002]{gas02}
  Gasteldello, F. \& Molendi, S. 2002, ApJ, 572, 160
\bibitem[1997]{gibson97} 
  Gibson, B.~K., Loewenstein, M., \& Mushotzky, R.~F.\ 1997, MNRAS, 290, 623
\bibitem[2001]{grandi01} 
  De Grandi, S. and Molendi, S., 2001, ApJ, 551, 153 
\bibitem[2001]{herder01} 
  den Herder, J.W., Brinkman, A.C., Kahn, S.M., et al. 2001, A\&A, 365, L7
\bibitem[2001]{jansen01}
  Jansen, F., Lumb, D., Altieri, B., et al. 2001, A\&A, 365, L1
\bibitem[1992]{kaastra92}
Kaastra, J.S. 1992, An X-Ray Spectral Code for Optically Thin Plasmas
(Internal SRON-Leiden Report, updated version 2.0)
\bibitem[2001]{kaastra01}
  Kaastra, J.S., Ferrigno, C., Tamura, T., et al. 2001, A\&A, 365, L99
\bibitem[2003a]{kaastra03a}
  Kaastra, J.S., Lieu, R., Tamura, T., Paerels, F.B.S., den Herder, J.W. 2003a, 
  A\&A, 397, 445
\bibitem[2004]{kaastra04}
  Kaastra, J.S., Tamura, T., Peterson, J.R., et al., 2004, A\&A, 413, 415
\bibitem[2003c]{kaastra03c}
  Kaastra, J.S., Lieu, R., Tamura, T., Paerels, F.B.S., den Herder, J.W. 2003c, 
  Adv. Sp. Res., in press.
\bibitem[1995]{liedahl95}
Liedahl, D.A., Osterheld, A.L., \& Goldstein, W.H. 1995, ApJL, 438, 115
\bibitem[2000]{matsushita00}
  Matsushita, K., Ohashi, T., \& Makishima, K.\ 2000, PASJ, 52, 685 
\bibitem[2003]{matsushita03}
  Matsushita, K., Finoguenov, A., \& B{\" o}hringer, H.\ 2003, A\&A, 401, 443
\bibitem[1985]{mewe85}
Mewe, R., Gronenschild, E.H.B.M., and van den Oord, G.H.J. 1985,
A\&AS, 62, 197
\bibitem[2001]{molendi01}
  Molendi, S., \& Gastaldello, F., 2001, A\&A, 375, L14
\bibitem[2003]{nevalainen03}
  Nevalainen, J., Lieu, R., Bonamente, M., \& Lumb, D., 2003, ApJ, 584, 716
\bibitem[2001]{peterson01}
  Peterson, J.R., Paerels, F.B.S., Kaastra, J.S., et al., 2001, A\&A, 365, L104
\bibitem[2003]{peterson03}
  Peterson, J.R., Kahn, S.M., Paerels, F.B.S., et al. 2003, ApJ, 590, 207
\bibitem[1993]{renzini93}
  Renzini, A., Ciotti, L., D'Ercole, A. \& Pellegrini, S., ApJ, 1993, 419
\bibitem[1992]{sarazin92}
  Sarazin, C.~L., O'Connell, R.~W., \& McNamara, B.~R.\ 1992, ApJL, 397, L31 
\bibitem[2002]{sakelliou02}
  Sakelliou, I., Peterson, J.R., Tamura, T., et al. 2002, A\&A, 391, 903
\bibitem[2001]{struder01}
  Str\"uder, L., Briel, U., Dennerl, K., et al. 2001, A\&A, 365, L18
\bibitem[2001a]{tamura01a}
  Tamura, T., Kaastra, J.S., Peterson, J.R., et al., 2001a, A\&A, 365, L93
\bibitem[2001b]{tamura01b}
  Tamura, T., Bleeker, J.A.M., Kaastra, J.S., et al. 2001b, A\&A 379, 107
\bibitem[2003b]{tamura03}
  Tamura, T., den Herder, J.W., \& Gonz\'alez-Riestra, R., 2003, 
XMM calibration note, XMM-SOC-CAL-TN-0034 (http://xmm.vilspa.esa.es/)
\bibitem[1995]{tsujimoto95} 
  Tsujimoto, T., Nomoto, K., Yoshii, Y., Hashimoto, M., 
  Yanagida, S., \& Thielemann, F.-K.\ 1995, MNRAS, 277, 945 
\bibitem[1991]{tsuru91} 
  Tsuru, T. 1991, PhD thesis, University of Tokyo
\bibitem[2001]{turner01}
  Turner, M.J., Abbey, A., Arnaud, M., et al. 2001, A\&A, 365, L27
\bibitem[2002]{xu02}
  Xu, H.~et al.\ 2002, ApJ, 579, 600 
\end{thebibliography}
\end{document}